\documentclass[aps,superscriptaddress,showpacs,floatfix]{revtex4}
\usepackage{epsfig,graphics,color}
\usepackage{graphicx}% Include figure files
\usepackage{dcolumn}% Align table columns on decimal point
\usepackage{bm}% bold math
\usepackage{eucal,eufrak}

\newcommand \tie {{\it i.e.}}

\newcommand \f {\not}

\newcommand \kd  {\delta}
\newcommand \ra  {\rightarrow}
\newcommand \imp {\Longrightarrow}
\newcommand \w  {\omega}

\newcommand \fp {{\bf p}}

\newcommand \fk {{\bf k}}
\newcommand \fq {{\bf q}}
\newcommand \fx {{\bf x}}
\newcommand \fy {{\bf y}}
\newcommand \fl {{\bf l}}
\newcommand \fP {\bm P}
\newcommand \h {\theta}

\newcommand \vk {\vec{k}}

\newcommand \vx {\vec{x}}
\newcommand \vp {\vec{p}}
\newcommand \vDl {\vec{\Delta}} 

\newcommand \mat {{\mathcal M}}
\newcommand \g {\gamma}
\newcommand \ro {\rho}
\newcommand \si {\sigma}
\newcommand \e {\epsilon}

\newcommand \p {{\prime}}

\newcommand \x {\cdot}
\newcommand \hf {\frac{1}{2}}
\newcommand \A {\alpha}
\newcommand \B {\beta}

\newcommand \lc {\langle}
\newcommand \rc {\rangle}

\newcommand \D {\Delta}
\newcommand \sg {\sigma}

\newcommand \nt {\noindent}
\newcommand \T {\tilde}

\newcommand {\llb} { \left[ \frac{\mbox{}}{\mbox{}} \right.}
\newcommand {\lrb} { \left. \frac{\mbox{}}{\mbox{}} \right] }

\newcommand \bvec{\left( \begin{array}{c} }
\newcommand \evec{\end{array} \right)}
\newcommand \tr {\mbox{{\bf Tr}}}
\newcommand \eg {{\it e.g.}}  
\newcommand \bea{\begin{eqnarray} }
\newcommand \eea{\end{eqnarray} }
\newcommand \nn {\nonumber}
\newcommand \be {\begin{equation}}
\newcommand \ee {\end{equation}}
\newcommand \epem {$e^+ e^-$}

\newcommand \sumint {\sum\!\!\!\!\!\!\!\!\int}
\newcommand \psibar {\bar{\psi}}
\newcommand \ata {& \times &}

\begin{document}

\title{Modification of the dihadron fragmentation function in nuclear matter}

\author{Abhijit Majumder}
\affiliation{Department of Physics, 
Duke University, Durham, NC 27708, USA.}

\author{Xin-Nian Wang}
\affiliation{Nuclear Science Division, 
Lawrence Berkeley National  Laboratory\\
1 Cyclotron road, Berkeley, CA 94720, USA.}

\date{ \today}

\begin{abstract} 
The medium modification of dihadron fragmentation functions in the deeply 
inelastic scattering (DIS) off a large nucleus is studied within the framework 
of the higher-twist expansion in the collinear factorization formalism. 
It is demonstrated that the
modification due to multiple parton scattering in the nuclear medium is similar 
to that of the single hadron fragmentation function. However, the conditional 
distribution of the associated hadron given by the ratio of dihadron 
to single hadron fragmentation function shows only slight modification.
The final results depend modestly on the nuclear density distribution. 
Comparisons with the experimental results on two hadron correlations 
as obtained by the HERMES collaboration at DESY are presented.
\end{abstract}


\maketitle

%%%%%%%%%%%%%%%%%%%%%%%%%%%%%%%%%%%%%%%%%%%%%%%%%%%%%%%%%%
%%%%%%%%%%%%%%%%%%%%%%%%%%%%%%%%%%%%%%%%%%%%%%%%%%%%%%%%%%
%%%%%%%%%%%%%%%%%%%%%%%%%%%%%%%%%%%%%%%%%%%%%%%%%%%%%%%%%%
%%%%%%%%%%%%%%%%%%%%%%%%%%%%%%%%%%%%%%%%%%%%%%%%%%%%%%%%%%
%%%%%%%%%%%%%%%%%%%%%%%%%%%%%%%%%%%%%%%%%%%%%%%%%%%%%%%%%%

\section{introduction}

%%%%%%%%%%%%%%%%%%%%%%%%%%%%%%%%%%%%%%%%%%%%%%%%%%%%%%%%%%
%%%%%%%%%%%%%%%%%%%%%%%%%%%%%%%%%%%%%%%%%%%%%%%%%%%%%%%%%%
%%%%%%%%%%%%%%%%%%%%%%%%%%%%%%%%%%%%%%%%%%%%%%%%%%%%%%%%%%
%%%%%%%%%%%%%%%%%%%%%%%%%%%%%%%%%%%%%%%%%%%%%%%%%%%%%%%%%%

The modification of the properties of jets as they pass through dense 
matter has emerged as a new diagnostic tool in the study of the partonic 
structure of such an environment \cite{Gyulassy:2003mc}. 
Such modification goes beyond a mere suppression of the leading hadron's 
multiplicity \cite{quenching,Baier:1996kr,guowang,Gyulassy:2000fs,Wiedemann:2000za,Turbide:2005fk} 
and could in principle be extended to include many particle
observables \cite{Feynman:1977yr,kon78,Fong:1990ph,maj04a,maj04d} such as
dihadron fragmentation functions. Defined as the two-hadron expectation 
values of partonic operators, these functions may be factorized \cite{col89,mue81} 
from the hard collisions and their evolution with the energy scale may be 
systematically studied in perturbative Quantum 
Chromodynamics (pQCD) \cite{maj04e,Majumder:2005vs}. As the single hadron 
fragmentation, dihadron fragmentation functions in vacuum can be defined
independently of the jet production processes. Once measured experimentally
as semi-inclusive distributions of di-hadrons from one hard process, they
can be used to predict dihadron distributions in other hard processes.
The medium modification of the dihadron fragmentation functions in the
multiple parton scattering picture will be dictated by jet-medium interaction
as in the single hadron modification and therefore will provide an
independent constraint on the jet transport parameters which in turn
can provide important information on the properties of the dense matter.

%The modification 
%thus involves a process dependent computation, which none the less many 
%be estimated in a controlled manner. The leading order and 
%leading twist computation is supplemented by contributions at leading logarithmic and 
%next-to-leading twist which receive a measure of support from the size of extended media 
%through which the jets propagate. 
%The isolation of these contributions involves 
%a knowledge of the color structure and parton density of the medium.  
%Thus, such observables hold great promise as deep probes of the medium 
%through which the partons propagate.

In both high-energy heavy-ion collisions and DIS off nuclei, two-hadron
correlations have been measured  \cite{adl03,Adler:2004zd,din04} in addition
to the single hadron spectra \cite{highpt,Airapetian:2000ks,Airapetian:2003mi}.
Two-hadron correlations are measured as the conditional probability 
distributions of a secondary hadron associated with a triggered hadron 
with given momentum. Therefore, they are given by the ratio of dihadron 
to single hadron fragmentation functions, while the single inclusive spectra 
are related to single hadron fragmentation function.
%Such measurements have recently been performed in 
%deep inelastic scattering off different nuclei by the HERMES experiment at DESY \cite{din04}.
%Measurements in hot and dense nuclear matter have been performed at the 
%Relativistic Heavy-Ion Collider (RHIC) by the STAR \cite{adl03} and PHENIX \cite{Adler:2004zd} 
%detectors.  
%The measured quantities differ between the two experiments and will 
%be addressed at some length in the upcoming sections of this paper. 
%Due to issues related to background subtraction and requisite statistics, 
%such measurements were restricted to two hadron correlations. 
%As a result, the theoretical description in the remainder of this paper 
%will be restricted to dihadron fragmentation functions. 
The study of the medium modification of single and dihadron fragmentation 
functions in heavy-ion collisions involves a somewhat different kinematics 
than that in the DIS experiments and will be discussed separately.
In this paper, the theoretical descripton of the dihadron correlations 
observed in DIS experiments off large nuclei will be presented.

In conformity with the conventional picture of DIS on nucleons,
the general process may be sketched as in Fig.~\ref{fig2}.
One of the quarks in the incoming nucleus receives a hard momentum 
transfer $\sim Q^2$ from the virtual photon. The struck 
quark then travels the remaning length of the nucleus 
while undergoing multiple scattering off soft 
gluons from various nucleons in its path. It escapes 
the nuclear environment and after some time, characterized 
by the confinement scale and dialated by its boost, undergoes 
the fragmentation process into a shower of hadrons.  
The presence of the hard scale ($Q^2$), provided by the momentum 
transfer between the lepton and the nucleus, allows for a 
factorized approach to the process of DIS off a nucleus, 
where the hard partonic parts can be calculated reliably within 
perturbative Quantum Chromodynamics (pQCD). 
At leading order, the differential hadronic cross section is 
estimated as a convolution of the quark distribution 
function in a nucleus with a leading order partonic cross section 
and the requisite fragmentation functions.
%
%None of the aforementioned 
%quantities involve a direct presence of the medium; with the sole 
%exception of the structure function which is enhanced by the atomic mass $A$.
%
The presence of the medium will be felt by the propagating quark through
multiple final state scattering at next-to-leading twist level and
the effects are in general suppressed by the hard scale $Q^2$,
%These are contained in diagrams of the type shown in Fig.~\ref{fig4}. 
While the final state interaction may be ignored in DIS off a nucleon, 
they play an important role in the DIS off a large nucleus where 
they are enhanced by the nuclear size $[A^{(1/3)}]$. This is due to 
the fact that the subsequent scattering with the gluon field 
(confined in nucleons) may occur at any of the nucleons which lie along
the path of the struck quark. Such medium effects can be factorised 
from the leading order hard cross section and manifest themselves 
as medium modified structure~\cite{lqs} and fragmentation 
functions~\cite{guowang} with the leading order (LO) hard partonic
part remains unchanged.

In this paper, we will focus our attention on the medium modification 
of the fragmentation functions. The nuclear modification of the quark
distribution functions will be included parametrically. We will work
within the formalism of higher-twist expansion in Ref.~\cite{lqs} to
calculate the nuclear enhanced power corrections to the semi-inclusive
cross section of DIS off a large nucleus. The modification of the single 
fragmentation functions within this formalism has been studied in Ref.~\cite{guowang}.
In Sec.~2 we give a brief review of the formalism and isolate the contibutions 
generic to the modification of the dihadron fragmentation functions.
The equations governing the medium modification of the dihadron 
fragmentation functions will be demonstrated to assume a form 
similar to that of the Dokshitzer-Gribov-Lipatov-Altarelli-Parisi (DGLAP)
evolution equations~\cite{gri72,dok77b,alt77} of the dihadron fragmentation 
functions in the vacuum~\cite{maj04a,maj04d}.
This is quite similar to the case of the modification of the 
single fragmentation functions. 
The additional term as compared to the DGLAP evolution equation will be 
a twist-four matrix element indentical to the case of the single 
fragmentation functions.
This matrix element will have to be evaluated in a nucleus of size $A$, and 
depends on the nuclear density distribution. 
In Sec.~3 the general methodology of the evaluation of this matrix element is 
outlined and its dependence on the nuclear density distribution is discussed.
Numerical results for the medium modification will be presented in Sec.~4
for two different nuclear density distributions and are compared with 
the experimental data from the HERMES experiment. We summarize and present 
our conclusions in Sec.~5.

%The exact 
%nature of the the measurements as well as a description of the 
%calculation will be presented in the following sections. 

%%%%%%%%%%%%%%%%%%%%%%%%%%%%%%%%%%%%%%%%%%%%%%%%%%%%%%%%%%
%%%%%%%%%%%%%%%%%%%%%%%%%%%%%%%%%%%%%%%%%%%%%%%%%%%%%%%%%%
%%%%%%%%%%%%%%%%%%%%%%%%%%%%%%%%%%%%%%%%%%%%%%%%%%%%%%%%%%
%%%%%%%%%%%%%%%%%%%%%%%%%%%%%%%%%%%%%%%%%%%%%%%%%%%%%%%%%%

\section{Formalism}

%%%%%%%%%%%%%%%%%%%%%%%%%%%%%%%%%%%%%%%%%%%%%%%%%%%%%%%%%%
%%%%%%%%%%%%%%%%%%%%%%%%%%%%%%%%%%%%%%%%%%%%%%%%%%%%%%%%%%
%%%%%%%%%%%%%%%%%%%%%%%%%%%%%%%%%%%%%%%%%%%%%%%%%%%%%%%%%%
%%%%%%%%%%%%%%%%%%%%%%%%%%%%%%%%%%%%%%%%%%%%%%%%%%%%%%%%%%

The focus of this article is restricted to the semi-inclusive 
process of DIS off a nucleus where hadrons are detected in the 
final state. In particular, we will consider the following 
process,

\bea
e(L_1) + A(p) \longrightarrow e(L_2) + h_1(p_1) + h_2(p_2) + X,
\label{chemical_eqn}
\eea

\nt
when at least two hadrons are detected in the final state,
where $L_1$ and $L_2$ represent the momentum of the 
incoming and outgoing leptons. The incoming nucleus of atomic mass 
$A$ is endowed with a momentum $Ap$. In the final state,
two hadrons ($h_1,h_2$) with momenta $p_1,p_2$ are detected.

\begin{figure}[htbp]
%\begin{center}
%  \epsfxsize 80mm
%\hspace{0cm}
  \resizebox{3in}{1.5in}{\includegraphics[0in,0in][8in,4in]{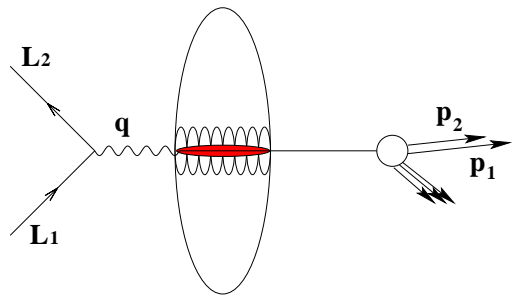}} 
%\vspace{0.25cm}
    \caption{ The lorentz frame of the process where a nucleon in a large nucleus is
    struck by a hard space-like photon.}
    \label{fig1}
%  \end{center}
\end{figure}

The kinematics is defined in a frame as sketched in Fig.~\ref{fig1}. 
In such a frame, the virtual photon $\g^*$ and the nucleus have 
momentum four vectors $\fq, {\bm P}_A$ given as, 

\[
\fq = {\bf L_2} - {\bf L_1} \equiv \left[\frac{-Q^2}{2q^-}, q^-, 0, 0\right], 
\mbox{\hspace{1cm}}
{\bm P}_A \equiv A[p^+,0,0,0],
\]

\nt
where we continue with the notation of Ref.~\cite{maj04a} of denoting four 
vectors in bold face. In this frame, the Bjorken variable is defined as 
$x_B = Q^2/2p^+q^-$. 

The differential cross section of the semi-inclusive process with two detected hadrons 
may be expressed as 

\bea
\frac{E_{L_2} E_{p_1} E_{p_2} d \sigma^{h_1,h_2}}{d^3 L_2 d^3 p_1 d^3 p_2} =
\frac{\A_{EM}^2}{2\pi s} \frac{E_{p_1} E_{p_2}}{ 4 Q^4}  L_{\mu \nu}  
\frac{d W^{\mu \nu}}{d^3 p_1 d^3 p_2},
\eea

\nt
where $s = (p+L_1)^2$ is the total invariant mass of the lepton nucleon 
system and the leptonic tensor is,

\bea 
L_{\mu \nu} = \tr [ \f L_1 \g_{\mu} \f L_2 \g_{\nu}].
\eea

\nt
In the notation used in this paper, $| A, p \rc$ represents the initial state of 
an incoming nucleus with $A$ nucleons each carrying momentum $p$. The 
hadronic part of the final state with two detected hadrons is represented as 
$| X, h_1 , h_2 \rc $.
As a result The semi-inclusive hadronic tensor may be defined as

\bea 
&& E_{p_1} E_{p_2} \frac{d W^{\mu \nu}}{d^3 p_1 d^3 p_2} = \sum_X 2\pi 
\kd (q + P_A - p_X -p_1 -p_2) \nn \\
&\times& \lc A,p |  J^{\mu}(0) | X, h_1 , h_2 \rc \lc X, h_1, h_2 | J^{\nu}(0) | A,p \rc ,
\eea

\nt
where the sum ($\sum_X$)  runs over all possible hadronic states and $J^{\mu}$ is the 
hadronic electromagnetic current ($J^{\mu} = e_q \bar{\psi}_q \g^\mu \psi_q$). It is 
understood that the quark wavefunctions are written in the Heisenberg picture. 
The leptonic tensor will not be discussed further. The focus in the remaining 
of this paper will be exclusively on the hadronic tensor. This tensor will be expanded 
order by order at the leading log. The leading and next-to-leading twist 
contributions that are nuclear enhanced will be isolated.

%%%%%%%%%%%%%%%%%%%%%%%%%%%%%%%%%%%%%%%%%%%%%%%%%%%%%%%%%%
%%%%%%%%%%%%%%%%%%%%%%%%%%%%%%%%%%%%%%%%%%%%%%%%%%%%%%%%%%
%%%%%%%%%%%%%%%%%%%%%%%%%%%%%%%%%%%%%%%%%%%%%%%%%%%%%%%%%%

\subsection{Leading twist}

%%%%%%%%%%%%%%%%%%%%%%%%%%%%%%%%%%%%%%%%%%%%%%%%%%%%%%%%%%
%%%%%%%%%%%%%%%%%%%%%%%%%%%%%%%%%%%%%%%%%%%%%%%%%%%%%%%%%%
%%%%%%%%%%%%%%%%%%%%%%%%%%%%%%%%%%%%%%%%%%%%%%%%%%%%%%%%%%

We start by noting that within the kinematics chosen there 
are two sets of final states in the opposite directions: one set $|S_1\rc$ is in the 
direction of the nuclear beam and consists mostly of its 
remnants, while the other $|S_2\rc$ is in the direction of the hard photon and 
consists of a jet of hadrons.
At leading twist, 
there is minimal overlap between these two sets and this leads to a simplified 
expression for the integrated hadronic tensor,

\bea
W^{\mu \nu} &=& \frac{1}{2} \sum_{\lambda_p} \sum_{h_1,h_2} 
\int \frac{d^3 p_1}{(2\pi)^3 2 E_1} 
\frac{d^3 p_2}{(2\pi)^3 2 E_2} \sum_{S_1 S_2} 
(2\pi)^4 
\kd^4 (\fq + \fP_A - {\bf p_{S_1}} - {\bf p_{S_2}} - {\bf p_{1}} - {\bf p_{2}})  \nn \\
&\times& \lc p \lambda_p | \bar{\psi}(0) | S_1 \rc \g^{\mu} \lc 0 | 
\psi(0) | S_2 p_1 p_2 \rc 
\lc S_2 p_1 p_2 | \bar{\psi}(0) | 0 \rc \g^\nu \lc S_1 | \psi(0) | p \lambda_p\rc. 
\eea

\nt
In the above equation, we include an average over the intial spin states $\lambda_p$ of the 
incoming nucleus, in the following this will be implied. We have also 
simplified $| A ; p \rc$ as simply $| p \rc$; the presense of $A$ should 
always be understood.
The above equation is restricted to the case of a quark inside the 
target undergoing a hard collision with the virtual photon. The case 
where an antiquark is struck is analogous and will not be discussed.  
The above is the leading order term in the usual expansion of the quark wavefunction 
operators in the 
interaction picture. In the rest of this section, 
the factorization of the dihadron fragmentation function from the 
hard part and the structure function will be discussed. It will be demonstrated that 
no new assumptions need to be invoked. 
%The factorization at next-to-leading order will 
%be discussed in the next subsection, where only the new  
The short hand notation 
\[
\sumint_{h_1 h_2} = \sum_{h_1,h_2} \int \frac{d^3 p_1}{(2\pi)^3 2 E_1} 
\frac{d^3 p_2}{(2\pi)^3 E_2}
\]
will be used henceforth. As in Ref.~\cite{maj04a} we use the notation that 
$p_1 +p_2 = p_h$.

The $\kd$-function can be espressed as an 
integral over four space-time, where phase factors involving the proton's momentum and that of its 
remnants may be absorbed into the matrix elements of the quark wavefunction operator 
on the proton state. 
Under a reorganization of the spinor matrix elements and introducing the 
momentum fractions $\zeta$, one obtains the hadronic tensor as 
 
\bea
W^{\mu \nu} &=& \sumint_{h_1 h_2} \int d^4 x e^{i \fx \x (\fq - \fp_{S_2} - \fp_h)} \int d^4 k d \zeta 
\kd^4 (\fp_{S_2} + \fp_h - \fk - \fq) \kd \left(\zeta - \frac{k^+}{p^+} \right) \nn \\
&\times& \tr \left[ \sum_{S_1} \lc S_1 | \psi (0) | p \rc \lc p | \bar{\psi} (x)  | S_1 \rc \g^\mu 
\lc 0 | \psi (0) | S_2 p_1 p_2 \rc \lc S_2 p_1 p_2 | \bar{\psi} (0) | 0 \rc \g^\nu \right] 
\label{l_t_reorg}
\eea 

\nt
In the above, the four-vector $k$ has been introduced which represents the momentum of the struck 
quark inside the nucleus. Using the four $\kd$-function, one may re-express 
$\exp[ i\fx \cdot (\fq - \fp_{S_2} - \fp_h)]$ 
as simply $\exp [-i\fx \cdot \fk]$. Taking the Fourier transform 
of the  $\kd$-function over $k$, leads to the 
expression,

\bea 
W^{\mu \nu} &=& \sumint_{h_1,h_2} \int d \zeta \int \frac{d^4 k}{ (2\pi)^4 } d^4 x  e^{-i \fx \cdot \fk } 
\kd (\zeta - \frac{k^+}{p^+}) \nn \\
&\times&\tr \left[ \lc S_1 | \psi(0) | p \rc \lc p | \bar{\psi} (x) | S_1 \rc \g^\mu \int d^4 y 
e^{-i \fy \cdot(\zeta \fp + \fq)}  
\lc 0 | \psi (0) | S_2 p_1 p_2 \rc \lc S_2 p_1 p_2 | \bar{\psi} (0) | 0 \rc \g^\nu \right] 
\eea

Invoking the collinear approximation that the 
($+$)-components of the target momenta $p^+,k^+$ are much larger than their ($\perp$)-components and 
($-$)-components, one may approximate the leading twist contribution of the $x$ dependent 
matrix element as 

\bea 
\hat{T}(\zeta,p) &=&  \int \frac{d^4 k}{ (2\pi)^4 } d^4 x  e^{-ix\cdot k} %
\kd (\zeta - \frac{k^+}{p^+}) 
\sum_{S_1} \lc S_1 | \psi(0) | p \rc \lc p | \bar{\psi} (x) | S_1 \rc  \simeq \frac{\f p}{2} f(\zeta) .
\eea

\nt
Where, $ f(\zeta) $ is identified as the unpolarized quark distribution function in a nucleus, 
which now has the obvious expression,
 
\bea
f(\zeta) &=& \int \frac{d^4 k}{ (2\pi)^4 } d^4 x  e^{-i \fx \cdot \fk} 
\kd (\zeta - \frac{k^+}{p^+}) 
\lc p | \bar{\psi} (x) \frac{\g^+}{2p^+} \psi(0) | p \rc  .
\eea

Thus, the initial quark distribution function may be factorized from the hard cross section 
and the final fragmentation into hadrons under the collinear approiximation. Factorization 
of the dihadron fragmentation function from the hard part requires the use of the collinear 
approximation in the opposite direction of the produced hadronic jet. One introduces the 
quark momentum in the direction of the hadronic jet as 
\[
\int d^4 l \kd^4 (\fl - (\zeta \fp + \fq)),
\]
as well as the momentum fractions of the two detected hadrons as 
\[
\int dz_1 dz_2 \kd \left(z_1 - \frac{p_1^-}{l^-}\right) \kd \left(z_2 - \frac{p_2^-}{l^-}\right).
\]
We continue with the notation of Ref.~\cite{maj04a} where $z = z_1 + z_2$.
As a result, the simplified expression for the differential hadronic tensor for the 
production of two hadrons with momentum fractions $z_1,z_2$ in the direction of the
quark jet is obtained as 

\bea 
\frac{W^{\mu \nu} }{dz_1 dz_2} &=& \int d\zeta \frac{f(\zeta) }{2\zeta} \sumint_{h_1,h_2}  
(2\pi)^4 \kd^4 \bigg( \frac{p_h}{z}  - q- \zeta p \bigg) \int \frac{d^4 l}{(2\pi)^4}  d^4 y 
e^{-iy\cdot l}\kd \bigg( z_1 - \frac{p_1^-}{l^-} \bigg) \bigg( z_2 - \frac{p_2^-}{l^-} \bigg) \nn \\ 
&\times& \tr \bigg[ \g^\nu (\frac{\f p_h}{z} - \f q) \g^\mu
\lc 0| \psi (0) | S_2 p_1 p_2 \rc \lc S_2 p_1 p_2 | \bar{\psi} (y) | 0 \rc \bigg].
\eea

\begin{figure}[htbp]
%\begin{center}
%  \epsfxsize 80mm
%\hspace{0cm}
  \resizebox{3.2in}{1.6in}{\includegraphics[0in,0in][8in,4in]{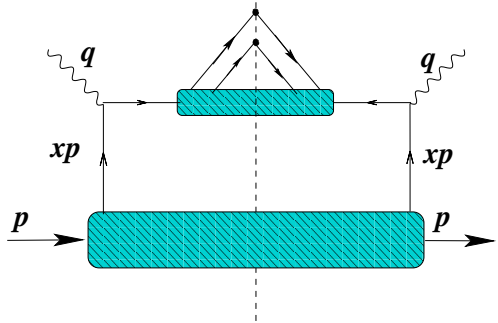}} 
%\vspace{0.25cm}
    \caption{ The Lowest order contribution to $W^{\mu \nu}$.}
    \label{fig2}
%  \end{center}
\end{figure}

The double integral over the momenta of the detected hadrons may be 
re-expressed as an integral over the 
sum of the two momenta and the difference. The functions $\kd ({p_h^-}{z}  - q^- - \zeta p^- ) 
\kd^2(\vec{p}_{h\perp})$ may be invoked to restrict the sum of the momenta of the detected hadrons. While 
in the limit afforded by the collinear approximation, the $\delta$-function over the (+)-componnets 
of the momenta yields, 
\[
\kd\bigg(\frac{p_h^+}{z} - q^+ - \zeta p^+\bigg) \simeq \kd \bigg(\frac{Q^2}{2q^-} - \zeta p^+\bigg) 
= \frac{\kd(\zeta - x_B)}{p^+}.
\]  
\nt
In the above, $x_B$ is the Bjorken scaling variable. The two $\delta$-functions over the two momentum fractions 
$z_1,z_2$ may be readjusted as demonstrated in Ref.~\cite{maj04a} which leads to the final factorized 
expression for the differential hadronic tensor as,

\bea
\frac{W^{\mu \nu} }{dz_1 dz_2} &=& \int d x_B f(x_B) H^{(0) \mu \nu} D_q^{h_1,h_2} (z_1, z_2), \label{LO_form}
\eea

\nt
where, $H^{(0)\mu \nu}$ is the hard part of the  quark-photon scattering
\tie, 

\bea 
H^{(0) \mu \nu} = \pi \kd [(q + x_Bp)^2 ] \tr [ \g^\nu \f p \g^\mu  (\f q - x_B \f p) ], \label{LO_fact}
\eea

\nt
and $D_q^{h_1,h_2} (z_1, z_2)$ is the dihadron fragmentation function,

\bea
D_q^{h_1,h_2}(z_1,z_2) &=& \frac{z^4}{4z_1z_2} 
\int \frac{d^2q_\perp}{4(2\pi)^3} 
\int \frac{d^4 l}{(2\pi)^4} \kd \left( z - \frac{p_h^-}{l^-}  \right)   
{\rm Tr} \Bigg[ \frac{\g^-}{2p_h^-} 
\int d^4 x e^{i \fp \x \fx} \sum_{S - 2} \nn \\
&\times& 
\lc 0 | \psi_q (x) | p_1, p_2, S-2 \rc  
\lc p_1, p_2, S-2 | \bar{\psi}_q (0) | 0 \rc \Bigg].
\eea

\nt
In the above expression, $q_\perp = p_{1\perp} - p_{2\perp}$. This expression is identical to that 
presented in Ref.~\cite{maj04a} with the obvious switch in the direction of the final momentum 
from the $(+)$ direction to the $(-)$ direction. Also, Eq.~(\ref{LO_fact}) may be directly generalized from 
the expression for single inclusive hadron production presented in Ref.~\cite{guowang}. 
The resulting picture in terms of Feynman diagrams is thus that of Fig.~\ref{fig2}. 
Hence, it may be stated that the factorization of the dihadron fragmentation from the 
LO diagram in DIS may be easily generalized from the factorization of the dihadron 
fragmentation function in \epem annihilation.

\begin{figure}[htbp]
%\begin{center}
%  \epsfxsize 80mm
%\hspace{0cm}
  \resizebox{3.2in}{1.6in}{\includegraphics[0in,0in][8in,4in]{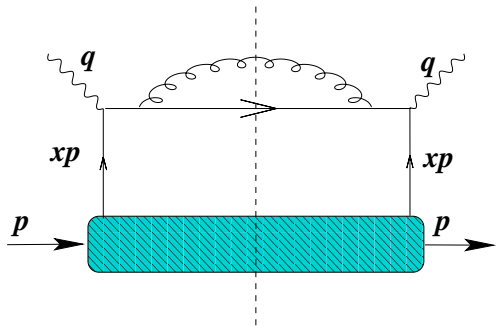}} 
%\vspace{0.25cm}
    \caption{ The hard part of the next-to-leading order contribution to $W^{\mu \nu}$.}
    \label{fig3}
%  \end{center}
\end{figure}

The method 
of factorization of the dihadron fragmentation function followed here may be easily repeated in 
the case at next-to-leading order to which we now generalize. The case at higher 
twist will be discussed in the next subsection. At next-leading-order in the strong coupling 
constant $\A_s$, the dominant contribution at the leading log level (in $n\cdot A=0$ gauge) 
comes from final 
state gluon radiation as shown in Fig.~\ref{fig3}. Specifically, these large logarithms result 
as the transverse momentum of the 
radiated gluon vanishes (\tie, $l_\perp \ra 0$).   
In this limit, the two detected hadrons may materialize in the fragmentation of the 
struck quark or from the radiated gluon. It is also possible that one hadron 
originates in the fragmentation of the quark while the other originates in the fragmentation of the 
gluon. In all these cases, the fragmentation functions may be factorised as a convolution with 
the hard cross section and a set of splitting functions, which encode the probability 
of a quark radiating a gluon which carries away a certain fraction of its forward momentum. 

No doubt, these contributions result in the 
Dokshitzer-Gribov-Lipatov-Altarelli-Parisi (DGLAP) evolution  of the dihadron fragmentation 
functions. These have been discussed at length in Refs.~\cite{maj04a,maj04d} for the 
case of jets in \epem collisions.
Using the methodology in these references and the preceding discussion at leading order, the leading  
log contributions to the hadronic tensor, as the radiated gluon becomes collinear to the struck 
quark, of the semi-inclusive DIS process may be expressed as 

\begin{widetext}
\bea
\frac{dW^{(1)}_{\mu \nu}}{dz_1 dz_2} &=& \sum_q \int dx f_q (x) H^{(0)}_{\mu \nu} 
\int_0^{\mu^2} \frac{dl_\perp^2}{l_\perp^2}\frac{\A_s}{2\pi} 
\Bigg[ \int_{z_1 + z_2}^1 \frac{dz}{z^2} P_{q\ra q g} (z)  
D_q^{h_1 h_2} \left( \frac{z_1}{z}, \frac{z_2}{z}\right) \nn \\
&+&  \int_{z_1}^{1-z_2} \frac{dz}{z(1-z)} \hat{P}_{q \ra q g}(z) 
D_{q}^{h_1} \left(\frac{z_1}{z} \right) 
D_g^{h_2} \left( \frac{z_2}{1-z} \right) \nn \\
&+& \int_{z_2}^{1-z_1} \frac{dz}{z(1-z)} \hat{P}_{q \ra q g}(z) 
D_{q}^{h_2} \left(\frac{z_2}{z} \right) 
D_g^{h_1} \left( \frac{z_1}{1-z} \right)\nn \\
&+& \int_{z_1 + z_2}^1 \frac{dz}{z^2} P_{q\ra g q} (z)  
D_g^{h_1 h_2} \left( \frac{z_1}{z}, \frac{z_2}{z} \right) \Bigg].
\label{q_dglap}
\eea
\end{widetext}

\nt
In the above equation, $H_{\mu \nu}^{(0)}$ represents the leading order hard cross section 
in Eq.~(\ref{LO_fact}). The quantities $P_{q\ra q g} (z)$  represent the 
probabilities that the struck quark radiated a gluon and retained a fraction $z$ of its energy. 
They have the well known expressions \cite{gri72,dok77b,alt77}, 

\bea
P_{q \ra qg} (z) = C_F \left( \frac{1+z^2}{1-z} \right)_+.
\eea

The subscript `+' indicates that the negative contribution from a virtual correction has been 
added within the splitting function. 
The splitting function in the second line of Eq. (\ref{q_dglap}),
$\hat{P}_{q\ra qg}$, is identical to the above equation except 
that it lacks the negative virtual correction. The other 
splitting function $P_{q\ra g q} (z)$ is obtained simply as $ P_{q\ra q g} (1-z)$ and 
also does not admit a virtual correction.

The gluon dihadron fragmentation function 
$D_g^{h_1 h_2} ( z_1, z_2 )$ is obtained as 

\bea 
D_g(z_1,z_2) &=& \frac{z^3_h}{2 z_1 z_2 }  \int \frac{dq_\perp^2}{8(2\pi)^2}
\int \frac{d^4l}{(2\pi)^4} 
\kd \left( z_h - \frac{p_h^-}{l^-} \right) \int d^4 x e^{i \fl \x \fx} \nn \\ 
&\times& \sum_{S-2} 
\lc 0 | A^a_\mu (x) | p_1,p_2,S-2 \rc 
\lc p_1,p_2,S-2 | A^b_\nu(0) | 0 \rc 
\frac{\kd^{ab} d^{\mu \nu}(l)}{16}, \label{glue_dihad}
\eea
where $d^{\mu \nu}(l)$ is the gluon's polarization tensor in the 
light-cone gauge and sum over the color indices of the gluon field 
is implied. The meaning of the various momenta and momentum fractions 
in Eq.~(\ref{glue_dihad}) is identical to the case of the quark 
fragmentation function. The single fragmentation functions $D_q(z),D_g(z)$ 
have the usual expressions \cite{col89, col82, mue78}. 

The NLO modification at leading twist presented above suggests the formulation 
of an effective next-leading-order and leading twist hard part. This includes  
the imaginary part of the photon nucleus forward scattering amplitude at order $\A_s$.
In particular, the hard part represents the process 
of a quark in the nucleus being struck by a hard photon into an off-shell intermediate 
state which then radiates a gluon, 

\bea
H^{(1)q}_{\mu\nu}(x,p,q,z)=H^{(0)}_{\mu\nu}(x,p,q) \int_0^{\mu^2} 
\frac{d\ell_T^2}{\ell_T^2} \frac{\alpha_s}{2\pi}
C_F \frac{1+z^2}{1-z} . \label{eq:h11}
\eea

\nt
This hard part is represented by the diagram in Fig.~\ref{fig3}. One can in principle 
deal solely with this hard part. The requisite double and single fragmentation functions 
may be convoluted with this hard part as a separate final step. 
This allows one to focus solely on the 
derivation of the hard part at the leading twist.  
Note that this hard part is identical to that derived for the case of the 
modification of the single fragmentation functions in Ref.~\cite{guowang}.
There remains the 
question as to whether such a separation may be achieved at higher twist. This is 
the subject of the next subsection.

%%%%%%%%%%%%%%%%%%%%%%%%%%%%%%%%%%%%%%%%%%%%%%%%%%%%%%%%%
%%%%%%%%%%%%%%%%%%%%%%%%%%%%%%%%%%%%%%%%%%%%%%%%%%%%%%%%%%
%%%%%%%%%%%%%%%%%%%%%%%%%%%%%%%%%%%%%%%%%%%%%%%%%%%%%%%%%%

\subsection{Higher Twist}

%%%%%%%%%%%%%%%%%%%%%%%%%%%%%%%%%%%%%%%%%%%%%%%%%%%%%%%%%%
%%%%%%%%%%%%%%%%%%%%%%%%%%%%%%%%%%%%%%%%%%%%%%%%%%%%%%%%%%
%%%%%%%%%%%%%%%%%%%%%%%%%%%%%%%%%%%%%%%%%%%%%%%%%%%%%%%%%%

In the calculation of cross sections which involve a hard scale, 
it often suffices to simply calculate the leading twist part. Higher twist contributions 
are suppressed by powers of the hard scale. It was demonstrated in 
Ref.~\cite{lqs}, that a class of higher twist operators are enhanced 
by the size of the target, $A^{1/3}$, where $A$ is the atomic mass. 
Specifically these contributions arise from the multiple scattering 
encountered by the struck quark off the soft gluons as it bore through the nucleus. 
Such contributions are also important in the medium modification of the single 
hadron fragmentation functions \cite{guowang}. In this section we 
isolate the hard part of the diagrams involved in the modification of the 
dihadron fragmentation functions.

\begin{figure}[htb!]
\begin{center}
%  \epsfxsize 80mm
%\hspace{-1cm}
  \resizebox{5in}{2in}{\includegraphics[1in,0in][9in,3in]{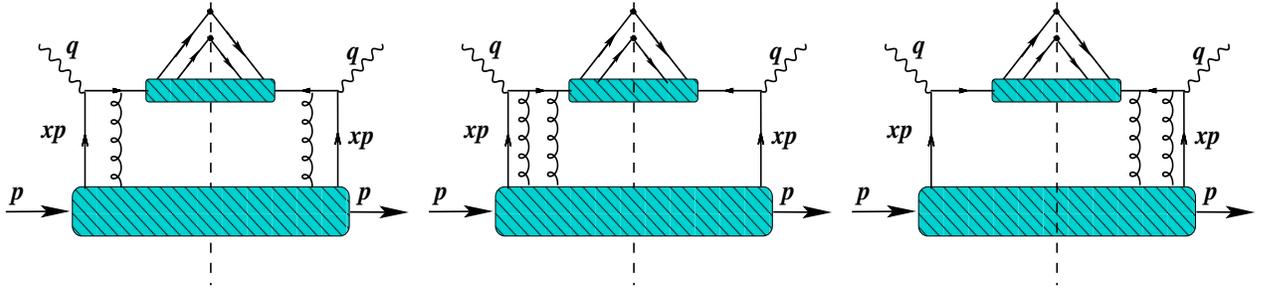}} 
%\vspace{0.25cm}
    \caption{ The leading order and next to leading twist contribution to $W^{\mu \nu}$.}
    \label{fig4}
\end{center}
\end{figure}

In the previous subsection, we demonstrated the factorization of the 
fragmentation and parton distribution functions from the hard scattering 
cross section at leading and next-to-leading order and at leading twist. 
It was demonstrated that the hard part is the same as in the case of the
single fragmentation functions. The focus will now be on a similar 
proof at next-to-leading twist. 
At leading order the diagrams to be considered are illustrated those in 
Fig.~\ref{fig4}.

The reader will immediately note that the procedure involved in the factorization of the 
dihadron fragmentation function from the hard part is entirely similar to that at leading 
twist, and also to the case of the single fragmentation functions. The resulting 
hard part has no radiated gluon and therefore no medium modification of the fragmentation 
functions. The sole effect of the double scattering indicated by the diagrams in Fig.~\ref{fig4}
is the appearence of a net transverse momentum $k_\perp$ in the out going jet brought in by the 
average transverse momentum of the gluons. In the limit of vanishing $k_\perp$, the contribution 
of the above diagrams is an overall Eikonal phase that will be absorded into a gauge invariant 
definintion of the quark distribution function.

\begin{figure}[htbp]
%\begin{center}
%  \epsfxsize 80mm
%\hspace{0cm}
  \resizebox{4in}{2in}{\includegraphics[1in,0in][9in,4in]{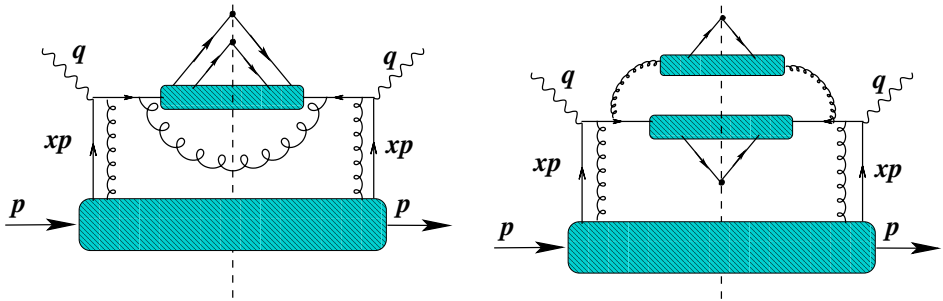}} 
%\vspace{0.25cm}
    \caption{ A contribution at next-to-leading order and next-to-leading twist to  $W^{\mu \nu}$.}
    \label{fig5}
%  \end{center}
\end{figure}

Medium modifications to the semi-inclusive cross section at higher twist in the limit of 
very small transverse momentum of the soft gluons in a nucleus emanate 
specifically at next to leading order. The contributions 
are illustrated by diagrams in Fig.~\ref{fig5}. Both diagrams represent 
the case where the struck quark undergoes scattering off a soft gluon in the nucleus and 
also radiates a fraction of its forward momentum into a collinear gluon. Both the quark and gluon 
then exit the medium and fragment. The diagram on the left corresponds to the case where both 
detected hadrons emanate from the fragmentation of the quark, while the diagram on the 
right corresponds to the case where one of the detected hadrons originates in the fragmentation 
of the quark and the other in the fragmentation of the gluon.
In Fig.~\ref{fig5} we two diagrams that arise at next-to-leading order and 
at next-to-leading twist in the evaluation of the medium modification of the 
dihadron fragmentation function. The dominant contributions, however, comes from
gluon-gluon secondary scattering \cite{guowang}.

As an example, we focus on the diagram on the right: 
the case of two independent fragmentations. The factorization of the 
dihadron fragmentation function from a quark line, as is the case for the diagram on the 
left in Fig.~\ref{fig5} is a simpler generalization from the case of the 
single inclusive cross section. Thus, the case of two single fragmentations 
represents the new addition to the type of diagrams involved in the calculation 
of the medium modification of the dihadron fragmentation functions. 
It will be demonstrated that 
the next-to-leading twist hard part, which arises from the evaluation of this contribution, 
is no different than 
the hard part, at next-to-leading twist, in the case of the single fragmentation functions. 
The difference lies in the convolutions involved: besides the usual convolution with the 
quark and gluon dihadron fragmentation functions, one also needs to convolute with a product of 
two single fragmentation functions as in Eq.~(\ref{q_dglap}).

As an illustraion, the contribution to the hadronic tensor from the 
right-hand diagram in Fig.~\ref{fig5} may written in general as

\bea
W^{\mu \nu} &=& e^2 g^4  t^d t^c t^a t^b \int d^4 y e^{i \fq \x \fy}  d^4 y_1 \sum_{S_1 S_2^a S_2^b} \sumint_{h_1 h_2} 
\lc p | \psibar (y) | Y \rc \sum_Y \lc Y | A_{\rho}^{c} (y_1) | S_1 \rc \g^\nu \lc 0 | A_\B^d (0) | S_2^b p_2 \rc  \nn \\
&\times& \lc S_2^b p_2 | A_\A^b (0) | 0 \rc \int \frac{d^4 k_2}{(2\pi)^4}  \frac{\f k_2 + \f l }{(k_2 + l)^2 - i \e} 
\g^\B \frac{\f k_2}{k_2^2 - i\e} 
e^{-i\fy \x (\fk_2 + \fl )} e^{-i \fy_1 \x (\fl_q - \fk_2) } \g^\rho \lc 0 | \psibar (0) | S_2^a p_1 \rc \nn \\
\ata\lc S_2^a p_1 | \psi(0) |0\rc \g^\si 
\int \frac{d^4 k}{(2\pi)^4} e^{ i\fy_2 \x (\fl_q - \fk)} \frac{\f k}{k^2 + i \e} \g^\A \frac{\f k + \f l }{(k+l)^2 + \e }
\g^\mu \lc S_1 | A_\si^a (y_2) | X \rc \sum_X \lc X | \psi(0) | p \rc .\label{h_t_1}
\eea

\nt
In the above equation, the gluon momentum $l = P_{S_2^b} + p_2$ and quark momentum $l_q = P_{S_1^a} + p_1$  are 
implicit. 
The sums $\sum_X, \sum_Y$ represent intermediate states assumed by the target after the initial quark is 
struck by the hard photon. The final state sums are $\sum_{S_1}$ in the direction of the target, $\sum_{S_2}$ in 
the direction of the photon. The latter is split into two parts $S_2^a$ emanating from the fragmentation of the 
quark and $S_2^b$ emanting from the fragmentation of the gluon.  The momenta $k,k_2$ represent the  momentum of the 
quark between the radiation of the gluon and scattering off the soft gluon field.  

We introduce the following factor of unity:
\[
\int d^4 l_q \int \frac{d^4 x_q}{(2\pi)^4} e^{i \fx_q \x (\fl_q - \fp_1 - \fp_{S_2^a})}  \\
\int d^4 l \int \frac{d^4 x_g}{(2\pi)^4} e^{i \fx_g \x (\fl - \fp_2 - \fp_{S_2^b})}  
\int d z_1^\p \kd \left( z_1^\p - \frac{p_1^-}{l_q^-} \right) \int d z_2^\p \kd \left( z_2^\p - \frac{p_2^-}{l^-} \right),
\]
followed by a reorganization of the spinor matrix structure as in Eq.~(\ref{l_t_reorg}) which leads to the 
trace over spinor indices. The introduction of the various integrals and $\delta$-functions allow for a simple 
procedure of indentifying and extracting the matrix elements in the fragmentation functions. Using 
the $\kd$-functions over the momentum fractions, one may re-express the partonic momenta $l_q,l$ in terms of the 
rescaled hadronic momenta $p_1/z_1^\p , p_2/z_2^\p$. As a result the quark fragmetation function may be isolated as,

\bea
D_q^{h_1} (z_1^\p) &=& \frac{{z_1^\p}^3}{2} \int \frac{d^4 l_q}{(2\pi)^4} \kd \left( z_1^\p - \frac{p_1^-}{l_q^-}  \right)
\int d^4 x_q e^{i \fx_q \x \fl_q }  \tr \left[ \frac{\g^-}{2p_1^-} 
\lc 0 | \psibar (x_q) | S_2^q p_1 \rc \lc S_2^b p_1 | \psi(0) |0\rc \right],
\eea   

\nt
and the gluon fragmentation function as, 

\bea 
D_g^{h_2} (z_2^\p) &=& \frac{{z_2^\p}^2}{2} \int \frac{d^4 l}{ (2 \pi)^4 } \kd \left( z_2^\p - \frac{p_2^-}{l^-}  \right)
\int d^4 x_g e^{i \fx_g \x \fl }    
\lc 0 | A^d_\B (x_g) | S_2^b p_2 \rc \lc S_2^b p_2 | A^b_\A (0) | 0 \rc d^{\A \B} \frac{\kd^{b d} }{8}.
\eea

\nt
Utilizing the above relations, one obtains the differential hadronic tensor as 

\bea
dW^{\mu,\nu} &=& dz_1^\p dz_2^\p e^2 g^4 \tr [t^d t^c t^a t^b]  \int d^4 y d^4 y_1 d^4 y_2 \sumint_{h_1,h_2} 
\tr \bigg[ \sum_{S_1} \lc S_1 | A^a_\sigma (y_2) \psi(0) | P \rc \lc P | \psibar(y) A^c_\rho (y_1) | S_1 \rc \g^\nu \nn \\
\ata \int \frac{d^4 k d^4 k_2}{(2\pi)^8}  \frac{\f k_2 + \f l}{(k_2 +l)^2 -i\e} \g^\B \frac{\f k_2}{k_2^2 - i\e }
\g^\rho \frac{\f p_1}{2} \g^\si \frac{\f k}{k^2 + i\e} \g^\A \frac{\f k + \f l}{(k +l)^2 + i\e} \g^\mu \bigg] 
d^{\A \B}\kd^{b d}
\nn \\
\ata e^{iy\x(q - k_2 - \frac{p_2}{z_2^\p})} e^{iy \x (k_2 - \frac{p_1}{z_1^\p})} e^{i y_2 \x (\frac{p_1}{z_1^\p} - k)}
\frac{D_q(z_1^\p)}{(z_1^\p)^3} \frac{D_g(z_2^\p)}{(z_2^\p)^2} .
\eea 

The factorization of the fragmentation functions from the 
hard part is completed by reexpressing the hadronic momentum 
integrals in terms of their rescaled partonic counterparts 
\tie, $d^3 p_1 \simeq d^3 l_q (z_1^\p)^3 $ and $d^3 p_2 \simeq  d^3 l (z_2^\p)^3$ . We also introduce the
variable $z$, the fraction of momentum that remains in the quark 
after it has radiated a gluon. In terms of $z$ the 
hadronic momenta are rescaled to $z_1 = z_1^\p z$ and 
$ z_2 = z_2^\p (1-z)$. Stipulating that the 
matrix elements of the initial states do not depend 
strongly on $y_\perp,{y_1}_\perp + {y_2}_\perp$, 
the integrals over the transverse coordinates may 
be completed yielding $\delta$-functions over the transverse momenta.
The matrix elements involving the initial states are dominated 
by the (-) component of their coordinates. 
Integrating over these transverse momenta under 
the influence of the above mentioned $\delta$-functions, 
the double differential hadronic tensor may be expressed as 

\bea 
\frac{d^2 W^{\mu \nu}}{dz_1 dz_2} &=& 
\int \frac{dz}{z(1-z)} D_q \left(\frac{z_1}{z}\right) 
D_g \left(\frac{z_2}{1-z}\right)
H^{(1)} (p,q,z)  \int dy^- dy_1^- dy_2^- d^2 y_\perp \nn \\ 
\ata \frac{1}{2}  e^{i y_\perp \x k_\perp} 
\lc P |  \psibar (y^-) \g^+  {A^a}^+ (y_1^-, y_\perp )  {A^a}^+ (y_2^-,0) \psi (0) | P \rc  \label{w_fact}.
\eea

\nt
The reader will note that the double differential cross section 
has been expressed as a convolution of an initial state  quark-gluon correlation  
in the incoming nucleon(nucleus), a hard scattering piece ($H^{(1)} (p,q,z)$) and 
final state quark and gluon fragmentation functions ($D_q(x),D_g(x)$). Isolation of 
the initial state piece involves the well known decomposition of the gluon vector potentials 
\cite{lqs},

\bea
A_\sg^a \simeq  \w_{\sg^\p, \sg} {A^a}^{\sg^\p}  +  p_\sg \frac{A^a \x n}{p \x n}, \label{A_breakup}
\eea

\nt
where, 

\[
\w_{\A, \B} = g_{\A \B}  -  \frac{ p_\A  n_\B  +  p_\B  n_\A }{ p \x n } .
\]

\nt
Within the given kinematics, the contribution from the second term in the Eq.~(\ref{A_breakup}) far out weighs 
that from the first term. This is followed by the usual approximation of isolating the leading twist piece of the 
quark gluon correlation function. The hard part in Eq.~(\ref{w_fact}) is obtained as

\bea 
H^{(1)} (p,q,z) &=&  g^4 \frac{\tr [t^d t^a t^a t^d]}{N_c^2-1} \tr \Bigg[ \frac{\f p}{2} \g^\nu 
\int \frac{d x_1 d x_2 d x_4  d k_\perp^2 }{ (2\pi)^5 }  \frac{d^4 l}{(2\pi)^4} \kd^+ (l^2) \kd^+ (l_q^2)
\frac{\f q + x_4 \f p}{ 2 q \x p  (x_B - x_4 + i\e)}   \nn \\
\ata  \g^\B \frac{\f q + x_4 \f p - \f l }{2 q \x p z (x_B - x_4 + x_L + i\e )}
\f p (\f q  + (x_1 + x_2) \f p - \f l + \f k_\perp )  \f p 
\frac{ \f q + x_1 \f p - \f l }{2 q \x p z (x_B - x_1 + x_L  - i\e )}  \nn \\
\ata \g^\A \frac{\f q  +  x_1 \f p }{2 q \x p ( x_B  - x_1 - i\e ) } \g^\mu  \Bigg] d_{\A \B}
\kd \left( 1 - z - \frac{l^-}{q^-} \right).
\eea   

\nt
In the above equation, $x_1,x_2,x_4$ represent the forward momentum fractions of the incoming quark and 
gluon and outgoing quark in Fig.~\ref{fig5}. The reader will also note that the hard part presented 
above in the evaluation of the double inclusive cross section is identical to the hard part 
in the evaluation of the single inclusive cross section \cite{guowang}. 

The remaining analysis 
of the cross section involves expanding the hard part as a Taylor expansion around the point $k_\perp = 0$.
The leading term in such an expansion yields the first term in the Eikonal expansion of the 
intermediate quark propagator in Fig.~\ref{fig3}. The second term which involves a linear derivative 
of the transverse momentum is zero identially for unpolarized targets. 
The third term survives and represents the first contibution 
to the nuclear enhanced higher twist contributions to semi-inclusive deep inelastic scattering. The 
evaluation of the higher twist contribution, factorized from the fragmentation functions, is 
in every way identical to that for the single inclusive cross section \cite{guowang}. 
Extracting the leading order hard part $H^{(0)}$ from $H^{(1)}$ and reorganizing the double inclusive 
cross section in the form of Eq.~(\ref{LO_form}), \tie,

\bea
\frac{dW^{\mu \nu} }{dz_1 dz_2} &=& \int d x_B f(x_B) H^{(0) \mu \nu} \tilde{D}_q^{h_1,h_2} (z_1, z_2), 
\eea

\nt
one obtains the medium modified dihadron fragmentation function  $\tilde{D}_q^{h_1,h_2} (z_1, z_2)$.

The modification of the fragmentation function in this sense depends on the scattering 
of the struck quark off the gluons in the medium and is thus dependent on the initial 
state. Following the methods employed in the computation of the medium 
modification of the single fragmentation functions, one obtains the 
medium modification of the dihadron fragmentation functions as the sum of 
higher twist contributions extracted from the sum of multiple diagrams at 
next-to-leading order. 
The presence 
of multiple fragmentation options allows for a greater number of contibutions to the 
equations governing the medium modification of the dihadron fragmentation functions. These may however be 
grouped together in a similar fashion to the vacuum evolution contributions to the dihadron fragmentation 
functions. 

\bea
\tilde{D}_q^{h_1,h_2} (z_1,z_2) &=& D_q^{h_1,h_2}(z_1,z_2) + 
\int_0^{Q^2} \frac{dl_{\perp}^2}{l_{\perp}^2} \frac{\A_s}{2\pi} \left[ \int_{z_h}^1 \frac{dz}{z^2} 
\left\{ \D P_{q\ra q g} (z,x_B,x_L,l_\perp^2) 
D_q^{h_1,h_2} \left(\frac{z_1}{z},\frac{z_2}{y} \right) \right. \right. \nn \\
&+&  
\left. \D P_{q\ra g q} (z,x_B,x_L,l_\perp^2) 
D_g^{h_1,h_2} \left(\frac{z_1}{z},\frac{z_2}{z} \right) \right\} 
+\int_{z_1}^{1-z_2} \frac{dz}{z(1-z)} \D \hat{P}_{q\ra q g} (z,x_B,x_L,l_\perp^2) \nn \\ 
&\times & \left. D_q^{h_1} 
\left(\frac{z_1}{z})\right) D_g^{h_2}\left(\frac{z_2}{1-z} \right)  
+ (h_1 \ra h_2) \right] \label{med_mod}
\eea

\nt
In the above $z_h=z_1+z_2$, and the switch $(h_1 \ra h_2)$ is only meant for the last term. 
$\D P_{q\ra q g},\D P_{q\ra g q}$ 
represent the medium modified splitting functions where a momentum fraction $z$ is left in 
the quark and the gluon respectively. 
Their expressions are identical to the modified splitting functions 
derived in Ref.~\cite{guowang}.
In the above, $x_L = l_\perp^2/(2p^+q^- z(1-z))$, where 
the radiated gluon or quark carries away a transverse momentum $l_\perp$.

As in the case for the evolution in the vacuum (Eq.~(\ref{q_dglap})), 
the splitting function $\hat{P}_{q\ra q g} = \mathfrak{Re}[P_{q \ra q g}]$ has no 
virtual counterpart and is given as 

\bea
\D \hat{P}_{q\ra q g} &=& \frac{1+z^2}{1-z}  \frac{C_A 2\pi \A   T^A_{qg} (x,x_L)}{(l_\perp^2 + 
\lc k_\perp^2 \rc)  N_c f_q^A(x,\mu_I^2)} .  \label{mod_split}
\eea

\nt
In the above equation, $C_A=3,N_c=3$. The scale $\mu_I^2$ 
represents the scale at which the quark distribution 
functions are factorized from the hard cross section 
at leading order and leading twist. The mean 
transverse momentum of the soft gluons is represented by the factor $\lc k_\perp^2 \rc$.  
The term $T^A_{qg}$ represents the quark gluon correlation in the
nuclear medium and includes contributions from multiple 
higher twist diagrams. The formal derivation  
of $T^A_{qg}$, identical to the case for the single 
fragmentation functions is outlined in Ref.~\cite{guowang}.
The final result which includes squares of amplitudes 
of soft-hard scatterings, hard double scattering and 
their interferences results in the simplified form,

\bea 
T^A_{q g} (x_B, x_L)  &=& \int \frac{dy^-}{2\pi} d y_1^-  d y_2^- e^{i (x_B + x_L) p^+ y^- } 
(1 - e^{-i x_L p^+ y_2^-}) (1 - e^{-i x_L p^+ (y^- - y_1^-) } ) \nn \\
\ata  \hf  \lc  p | \psibar (0)  \g^+ F_\sg^+ ( y_2^- ) {F^{+}}^\sg (y_1^- ) \psi (y^-) 
| p \rc  \h (-y^-) \h (y_2^- - y_1^- ). \label{mod_factor}
\eea

\nt 
Up to this point the state $|p \rc$ was generically referred to as that of a 
nucleus with momentum $Ap$.
Indeed, the isolation of the higher twist piece in DIS so far has been independent of the 
content of this state. If $|p \rc$ represented a nucleon with momentum $p$ then the above equation 
would represent the medium modification factor for DIS off a nucleon. The enhancement of 
such higher twist objects in extended nuclear media will be the subject of the subsequent sections.
%In the following section, the state $| p \rc$ will specifically mean that of 
%nucleus with total momentum $Ap$ where $A$ is the atomic mass of this nucleus.  
This will entail  the decomposition of the nuclear medium into nucleons and will directly involve 
the nuclear density 
distribution. The evaluation of $T^A_{qg}$ will expose the eventual nuclear size 
enhancement and as a result justify the incorporation of the higher twist contributions in the 
evaluation of inclusive hadron production in the DIS off a large nucleus.

%%%%%%%%%%%%%%%%%%%%%%%%%%%%%%%%%%%%%%%%%%%%%%
%%%%%%%%%%%%%%%%%%%%%%%%%%%%%%%%%%%%%%%%%%%%%%
%%%%%%%%%%%%%%%%%%%%%%%%%%%%%%%%%%%%%%%%%%%%%%
%%%%%%%%%%%%%%%%%%%%%%%%%%%%%%%%%%%%%%%%%%%%%%
%%%%%%%%%%%%%%%%%%%%%%%%%%%%%%%%%%%%%%%%%%%%%%

\section{Evaluation of the Nuclear Modification factor}

%%%%%%%%%%%%%%%%%%%%%%%%%%%%%%%%%%%%%%%%%%%%%%
%%%%%%%%%%%%%%%%%%%%%%%%%%%%%%%%%%%%%%%%%%%%%%
%%%%%%%%%%%%%%%%%%%%%%%%%%%%%%%%%%%%%%%%%%%%%%
%%%%%%%%%%%%%%%%%%%%%%%%%%%%%%%%%%%%%%%%%%%%%%
%%%%%%%%%%%%%%%%%%%%%%%%%%%%%%%%%%%%%%%%%%%%%%

In the preceding section, the higher-twist contribution to the inclusive 
hadron production in deep-inelastic scattering off a nucleus was expressed 
as the convolution of the quark distribution function in a nucleus, a hard LO photon-quark scattering 
cross section and a nuclear modified fragmentation function. The equations 
governing the modification of the dihadron fragmentation function in medium 
(Eq.~(\ref{med_mod})) depend on the nuclear modification factor (Eq.~(\ref{mod_factor})). This 
quantity essentially represents a quark-gluon correlation function within the nucleus. 
The evaluation of this correlation function is the subject of this section.

The essential kinematics for the process of DIS off a large nucleus has been 
outlined in Sec.~II. A virtual photon with momentum $\fq$ strikes a 
large nucleus with atomic mass $A$ with a momentum $A\fp$. 
In the evaluation of the multiple scattering of the struck quark in a 
large nucleus we will invoke the convolution model, which decomposes the 
deep-inelastic scattering off a nucleus in terms of the scattering off its 
constitutient nucleons. In this paper, we will follow the version of the 
convolution model outlined in Ref.~\cite{Osborne:2002st}.
The evaluation of the quark-gluon correlation function which arises as the struck 
quark traverses such a  nucleus commences with the 
decomposition of the nuclear ket $| p; A \rc$ in terms of nucleons,

\bea 
| p; A \rc = \int \prod_{i = 1}^A  \frac{ d^3 p_i  \h( p_i )}{ (2\pi)^3  2 p_i^+  } 
\Phi (\{ p_i \}) | \{ p_i \} \rc   (2\pi)^3  2 p^+ \kd ^3 (\sum_i \vec{p}_i -  A p ). \label{nuc_state}
\eea

\nt
In the above equation, $\Phi (\{ p_i \})$ represents the nuclear wavefunction; the ket 
$| \{ p_i \} \rc  = | p_1, p_2, \ldots, p_A \rc $ represents a particular ket where the 
nucleons $1 \ldots A$ assume momenta $p_1, \ldots ,  p_A$. The overall three momentum conservation 
is enforced via the $\delta$-function. In the notation employed in this article, a single nucleon 
state is normalized as 

\bea 
\lc p_i |  p_j \rc = 2p_i^+ (2 \pi)^3 \kd^3 (\vec{p}_i - \vec{p}_j) ,  \label{mom_delta}
\eea

\nt
as a result the $n$-nucleon state is normalized as 

\bea 
\lc \{ p_i \} | \{ p_j \} \rc = \prod_{i=1}^n  2 p_i^+ (2 \pi)^3 \kd^3 ( \vec{p}_i  -  \vec{p}_j ).
\eea

\nt
Defining the nuclear momentum $P_A = A p$, the normalization of the nuclear state may be 
stipulated as,

\bea 
\lc p ; A | p^\p ; A \rc =  2 p^+ (2 \pi)^3 \kd^3 ( \vec{P}_A  -  \vec{P^\p}_{A} ) . \label{nuc_norm}
\eea

\nt
Substituting the expression for $| p ; A  \rc$ from Eq.~(\ref{nuc_state}) into Eq.~(\ref{nuc_norm}) 
leads to the normalization condition on the nuclear wavefunction (where we use the simplified notation 
$ d \prod_{1}^A = \prod_{i=1}^A d^3 p_i \h (p_i^+) / [(2\pi)^3 2p_i^+ ]$):

\bea 
\int d \prod_1^A  \Phi^* (\{ p_i \}) \Phi (\{ p_i \}) (2 \pi)^3 2 p^+ \kd^3 (\sum_i \vec{p}_i  - \vec{P}_A) = 1.
\eea

Halting one of the momentum integrations on the {\it l.h.s.} of the above equation  leads to the momentum 
space one-nucleon density, which represents the probability of finding a nucleon with momentum $k$, 

\bea
\rho(k) = \int d \prod_{2}^{A}  | \Phi (k, \{ p_i \})|^2 (2  \pi)^3 2 p^+ 
\kd \Big( Ap^+ - k^+ - \sum_{i=2}^A p_i \Big)
\kd^2 \Big( \vec{k}_\perp  -  \sum_2^A \vec{p}_{i \perp}  \Big),
\eea

\nt
where, we have decomposed the 3-$\kd$ function into its longitudinal and transverse components.
Halting two integrations leads to the two-nucleon momentum correlator $\rho(k_1,k_2,\D)$,
which contains information regarding the sharing of momentum between nucleons in a nucleus,

\bea
\rho(k_1,k_2,\D) &=& \int d \prod_{3}^{A} \Phi^* (k_1 + \D/2, k_2 - \D/2 , \{ p_i \})  
\Phi (k_1 - \D/2, k_2 + \D/2 , \{ p_i \})  \nn \\
\ata (2  \pi)^3 2 p^+ \kd \Big( Ap^+ - k_1^+ - k_2^+ - \sum_{i=3}^A p_i \Big)
\kd^2 \Big( \vec{k}_{1,\perp}  +  \vec{k}_{2,\perp} -  \sum_{i=3}^A \vec{p}_{i \perp}  \Big).
\eea

\nt
Both these and the higher correlations obey the obvious sum rules, \eg,

\bea 
\int \tilde{d k} \rho(k) = 1,  \mbox{\hspace{1cm}}  
\int \tilde{d k}_1 \tilde{d k}_2  \rho(k_1,k_2,0) = 1,
\eea

\nt
where we have used the  short hand $\tilde{d k} = d^3 k \h (k^+) / [(2\pi)^3 2k^+ ]$.

The 
matrix element included in the definition of the modification factor in Eq.~(\ref{mod_factor}) 
involves interferences between the different processes which contribute to different phase factors in 
Eq.~(\ref{mod_factor}). The interference between the various processes is thus dependent on these 
factors and ultimately on the locations $y,y_1,y_2$ where the various scatterings have taken place. 
Hence, the nuclear wavefunctions discussed above need to be transformed to position space. In the 
notation adopted in this paper, the position bases demonstrate the following outer product and 
normalization,

\bea  
\int \prod_i d^3 x_i | \{ x_i \} \rc \lc  \{ x_i \} | = 1 , \mbox{\hspace{1cm}} 
\lc x_i | p_i \rc = e^{i \vp_i \x \vx_i} \sqrt{2 p_i^+} . \label{op_norm}
\eea

\nt
Within this basis, the nuclear ket may be expressed as  
$| A \rc   =  \int  \prod_i  d^3 x_i  \Phi({x_i})  | \{ x_i \} \rc$.
To simplify the expessions, we use the over all $\delta$-function 
in Eq.~(\ref{nuc_state}) to carry out 
the integration of the $A^{th}$ momentum. There are now simply $A-1$ nucleon 
momenta to be dertermined to isolate as kiven ket, the wavefunction provides 
the probability amplitude for a given choice of $A-1$ momenta such that 
one of them is specified.
The momentum of the final ket $p_A$ is such that the momentum of the nucleus 
remains $P_A$ \tie, $p_A^+ = P_A^+ - \sum_i^{A-1} p_i^+$ . 
Introducing a complete basis of position states on the {\it r.h.s.} 
of Eq.~(\ref{nuc_state}), one can reexpress the nuclear state in position space
as

\bea
| p ; A \rc = \int  \prod_{i=1}^{A-1} d^3 x_i \int  
\prod_{i=1}^{A-1} \tilde{d p}_i \frac{ \h (p_A^+)}{\sqrt{2 p_A^+}}
\Phi (\{ p_i \}, p_A) e^{i \sum_i^{A} p_i \x x_i } 2p^+ | \{ x_i \}  \rc . 
\eea

\nt
By mere inspection, the nuclear wave function in position space may 
be expressed as

\bea 
\Phi( \{ x_i \}) = \int \prod_i^{A-1} \tilde{d p}_i \frac{ \h (p_A^+)}{\sqrt{2 p_A^+}} 
\Phi( {p_i}, p_A)  e^{i\sum_i^{A} \vp_i \x \vx_i} 2p^+,
\eea

\nt
where, once again, it is understood that $p_A = P_A - \sum_{i=1}^{A-1} p_i$.  As a result 
the wavefunction in position space has the right translational properties, \tie,
\[ \Phi(\{ x_i + \D x \})  = e^{i P_A \x \D x} \Phi (\{ x_i \}).\] 
The inverse transform is obtained as 

\bea
\Phi(\{p_i\}) = \frac{1}{p^+} \int \prod_{i=1}^{A-1} d^3 x_i \sqrt{2 p_i^+} 
e^{-i \sum_{i=1}^{A-1} \vp_i \x \vx_i  } \Phi(\{ x_i \}) \sqrt{2P_A^+ - 2 \sum_i^{A-1} p_i^+}. 
\eea

Armed with the decompositions above for the nuclear wavefunction, 
one can now evaluatr the nuclear modification factor in  Eq.~(\ref{mod_factor}).
The primary object to be evaluated in Eq.~(\ref{mod_factor}) is the matrix element 
\[
\mat = \lc p; A | \psibar (0)  \g^+ F_\sg^+ ( y_2^- ) {F^{+}}^\sg (y_1^- ) \psi (y^-) 
| p ; A \rc .  
\]
The quark and gluon operators carry color. However, the expectation of this operator 
is evaluated in a nucleus which is color neutral. In the convolution model 
outlined above the nucleus is decomposed into its constituent color neutral nucleons.
The various position variables $0,y^-,y_1^-,y_2^-$ represent the locations in the 
nucleus where the quark was struck by the hard photon and where the soft gluon was 
encountered by the struck quark. As the quark and gluon carry color in different 
representations of SU(3), both quark operators must be confined within the same nucleon, as 
must both gluon operators. It is possible that all four operators are confined within 
the same nucleon. This is definitely the case for the higher twist contribution to DIS off 
a nucleon. In a nucleus, the two quarks operators and the two gluon operators may 
be confined to two different nucleons. In the limit in which the struck quark carries 
a large forward momentum $x_1 p^+$ and the soft gluon carries a very small transverse momentum 
$k_\perp$, the struck quark follows an almost linear trajectory as it burrows through 
the nucleus. As a result the struck gluon may be located in any of the nucleons 
that lie directly behind the nucleon which is struck by the photon. This leads to an 
enhancement of the order of one nuclear dimension \tie, $\sim A^{1/3}$. In this and 
the next section, this factor of $A^{1/3}$ will be explicitly extracted. This 
nuclear size enhancement, {\it aposteriori}, justifies the following approximation for the 
matrix element, 

\bea 
\mat &=& {A \choose 2} \int \prod_{i=3}^{A-1} \tilde{dp}_i  \frac{d^3 k_1  d^3 k_2 d^3 \D 
\h (k_1^+ + \D^+/2) \h (k_1^+ - \D^+/2) \h (k_2^+ + \D^+/2) \h (k_2^+ - \D^+/2 ) }
{ (2\pi)^9 2^4 ( {k_1^+}^2  -  {\D^+}^2/4  )  (  {k_2^+}^2 -  {\D^+}^2/4 ) }
\frac{\h(p_A^+)}{2 p_A^+} \nn \\
\ata \Phi^* (k_1 - \D/2 , k_2 + \D/2 , \{p_i\}_3^{A-1} )
\Phi (k_1 + \D/2 , k_2 - \D/2 , \{p_i\}_3^{A-1} )   (2p^+)^2  \nn \\
\ata \lc k_1 - \D/2 | \psibar(0) \g^+ \psi(y^-) | k_1 + \D/2 \rc
\lc k_2 + \D/2 | F_\sg^+  (y_2^-)  F^{\sg +} (y_1^-)  | k_2 - \D/2 \rc . \label{nucleon_mat}
\eea

\nt
In the above equation, we have simply reexpressed the nuclear ket in terms of  
nucleon kets in momentum space as required by Eq.~(\ref{nuc_state}). 
The set of states $\{p_i\}_3^{A-1}$ represents $3$ to $A-1$ nucleons, 
the momenta of the first two nucleons are $k_1 \pm \D/2$ and $k_2 \pm \D/2$.
Due to color confinement the two quark operators in $\mat$ act on the same nucleon, 
as do the two gluon operators. The factor 
of ${A \choose 2}$ originates in the number of ways the two nucleons may be chosen.
The remaining $A-2$ nucleons are unaffected by the operators and lead to the appearence 
of $A-2$ momentum $\delta$-functions as in Eq.~(\ref{mom_delta}). 
Integrating over these 
momenta removes $A-2$ of the $2A -2 $ integrals leaving the $A$ integrals as in the 
equation above. 
The momenta of the two nucleons on the {\it r.h.s.} of the Eq.~(\ref{nucleon_mat})
which form the bras and kets of the two partonic operators are such that the 
sum of the momenta of 
the two nucleons that form the bras are exactly equal to the sum of the momenta of the two 
nucleons that form the kets and each is equal to $k_1+k_2$. This mismatch in 
momentum is the reason for the appearence of the two off-forward parton distributions~\cite{Ji:1996ek} which 
will result from the fourier transform of the following shifted matrix elements of the partonic operators,

\[
\lc k_1 - \D/2 | \psibar(0) \g^+ \psi(y^-) | k_1 + \D/2 \rc
\lc k_2 + \D/2 | F_\sg^+  (y_2^-)  F^{\sg +} (y_1^-)  | k_2 - \D/2 \rc. 
\]

Utilizing the expressions afforded by the preceding discussion, the matrix element in 
Eq.~(\ref{nucleon_mat}) may be transformed to position space. 
The $A-2$ $\h$-functions which contain all but the mismatched momenta 
\tie \, $p_3^+, \ldots p_A^+$ are Fourier transformed,

\bea 
\h (p_i^+) = \int \frac{d z_i}{2\pi i} \frac{ e^{i z_i p_i^+} }{ z_i - i \e }. \label{theta}
\eea

\nt 
The $A-2$ momentum integrations can now be carried out setting the position of the 
i$^{th}$ nucleon in $\Phi^*$  to $x_i^\p = x_i + z_i$. Utilizing the fact that 
the nuclear wavefunctions are rapidly dropping functions as any one of the coordinates 
tends to infinity allows the evaluation of the $A-2$ countour integrations 
over the variables $z_i$, yielding the expression 

\bea 
\mat &=& {A \choose 2} \int \frac{d^3 k_1  d^3 k_2 d^3 \D 
\h (k_1^+ + \D^+/2) \h (k_1^+ - \D^+/2) \h (k_2^+ + \D^+/2) \h (k_2^+ - \D^+/2 ) }
{ (2\pi)^9 2^2 \sqrt{( {k_1^+}^2  -  {\D^+}^2/4  )  (  {k_2^+}^2 -  {\D^+}^2/4 )} } \nn \\
\ata \int d^3 x_1 d^3 x_2 d^3 x_1^\p d^3 x_2^\p  \Phi^* (x_1^\p,x_2^\p, \{x_i\}) 
\Phi (x_1,x_2,\{x_i\}) \nn \\ 
\ata \exp \bigg[ -i(\vk_1 + \vDl/2 )\x \fx_1   -i(\vk_2 - \vDl/2 )\x \fx_2  
+i(\vk_1 - \vDl/2 )\x \fx_1^\p   + i(\vk_2 - \vDl/2 )\x \fx_2^\p  \bigg] \nn \\
\ata \lc k_1 - \D/2 | \psibar(0) \g^+ \psi(y^-) | k_1 + \D/2 \rc
\lc k_2 + \D/2 | F_\sg^+  (y_2^-)  F^{\sg +} (y_1^-)  | k_2 - \D/2 \rc . \label{nucleon_mat_x}
\eea

>From Eq.~(\ref{mod_factor}), the complete nuclear modification factor is composed of 
four terms, each of which contains the matrix element expressed in Eq.~(\ref{nucleon_mat_x}).
The four terms represent the squares of amplitudes and the interferences between double 
hard and hard soft scattering. Each of these terms may be generically expressed as 

\bea
T &=& \int \frac{dy^-}{2 \pi} d y_1^- d y_2^- e^{i x_A p^+ y^-  + i x_B p^+ (y_1^- - y_2^-) 
+ i x_C p^+ y_2^- } \frac{1}{2} \mat (y^-,y_1^-,y_2^-) \h(-y_2^-)  \h( y^- - y_1^- ), \label{general_term}
\eea

\nt
where, $x_A, x_B, x_C$ represent different combinations of momentum fractions, and 
$\mat (y^-,y_1^-,y_2^-)$ represents the same matrix elements as those of 
Eqs.~(\ref{nucleon_mat}) and (\ref{nucleon_mat_x}) with the dependence on the partonic 
position variables shown explicitly. As the reader will note, the full form of 
$T$ is essentially a convolution between two factors one that contains an 
integral over nucleon position variables 
and another that contains an integral over partonic position 
variables. These two terms are convoluted through the nucleon momenta $k_1,k_2,\D$  \tie

\bea
T = \int d^3 k_1 d^3 k_2 d^3 \D  \big[ N(\vk_1,\vk_2,\vDl ) P(\vk_1,\vk_2,\vDl )  \big] ,
\eea
where,
\bea 
N(\vk_1,\vk_2,\vDl ) = \int d^3x_1 d^3 x_2 d^3 x_1^\p d^3 x_2^\p 
\mathcal{N}(\vk_1,\vk_2,\vDl, \vx_1,\vx_2,\vx_1^\p,\vx_2^\p ) 
\eea
and 
\bea 
P(\vk_1,\vk_2,\vDl) = \int d y^- d y_1^- d y_2^- 
\mathcal{P} (\vk_1,\vk_2,\vDl, y^-,y_1^- , y_2^-).
\eea

New partonic position variables with unit Jacobian $u^- =  y_1^- - y_2^-$ 
and $w^- = (y_1^-  + y_2^- )/2 $ are introduced. In terms of these the 
partonic matrix elements of the gluon operators may be simplified as 

\bea
\lc k_2 + \D/2 | F_\sg^+  (y_2^-)  F^{\sg +} (y_1^-)  | k_2 - \D/2 \rc
= \lc k_2 + \D/2 | F_\sg^+  \left( \frac{-u^-}{2} \right)  F^{\sg +}
 \left( \frac{u^-}{2} \right)  | k_2 - \D/2 \rc e^{i \D^+ w^- }.
\eea 

\nt
Similarly, the $\h$-functions over the partonic locations may be reexpressed as

\bea
\h(-y_2^-) \h (y^- - y_1^-) = \h(u^-/2 - w^-) \h (y^- - w^- - u^-/2) = 
\int \frac{dk dl }{(2\pi i)^2}  \frac{e^{-ik ( w^- - u^-/2 )}}{ k - i\e }
\frac{e^{-il (y^- - w^- - u^-/2 )}}{ l - i\e } .
\eea 

\nt
Integrating out $w^-$ one obtains the constraint $2\pi \kd (\D^+  - (k + l - x_C p^+))$.
This $\delta$-function is used to perform the integral over the longitudinal 
off-set in the 
momenta \tie, $\D^+$. It has already been stated that the matrix elements of the 
partonic operators in Eqs.~(\ref{nucleon_mat},\ref{nucleon_mat_x}) are, under the 
action of Fourier transforms, related to the off-forward parton distributions (OFPD) 
\cite{Ji:1996ek}. However, as was pointed out in Ref.~\cite{Ji:1997gs}, MIT bag models 
of the OFPD's suggest that the variation with $\vDl$ is rather slow and thus 
we will assume that the OFPD's remain more or less constant with $\vDl$. It will also 
be assumed that the matrix elements are almost independent of the transverse 
momentum of the nucleons \tie, $k_{1 \perp}, k_{2 \perp}$. This allows 
the integration over $k_{1 \perp}, k_{2 \perp}$ and $\D_\perp$ which leads to three 
sets of $\delta$-functions over the transverse components of the nucleons involved: 
$\kd^2 (\vx_{1 \perp} - \vx^\p_{1 \perp})$ , $\kd^2 (\vx_{2 \perp} - \vx^\p_{2 \perp})$ 
and  $\kd^2 (\vx_{1 \perp} - \vx_{2 \perp})$. Thus all the transverse positions of the 
nucleons are set to be the same. As a result, the simple linear trajectory of the 
propagation of the struck quark as presented in Secs.~II \& III is reinstated. The 
representative matrix element, with the incorporation of the above simplifications, 
assumes the form

\bea
T &=& \int \frac{ dk dl}{ (2 \pi i)^2}  \int \frac{d k_1^+ d k_2^+}{(2 \pi)^2} 
\frac{ \h(k_1^+ + \D^+/2 )  \h (k_1^+ - \D^+/2) \h (k_2^+ + \D^+/2)  \h (k_2^+ - \D^+/2 ) }
{ 4 \sqrt{( {k_1^+}^2  -  {\D^+}^2/4  )  (  {k_2^+}^2 -  {\D^+}^2/4 )} } \nn \\
\ata {A \choose 2} \int d x_1^- d x_2^- d \kd_1 d \kd_2  d^2 x_\perp \prod_{i=3}^{A-1} d^3 x_i  
e^{i k_1 \kd_1 } e^{ i k_2 \kd_2} e^{i \D^+ ( x_2^- - x_1^- )} \nn \\
\ata \Phi^* (\{ x_1^- + \kd^-_1/2, x_\perp \} , \{ x^-_2 + \kd^-_2/2 , x_\perp \} , \ldots ) 
\Phi ( \{ x_1^- - \kd^-_1/2 , x_\perp \}, \{ x^-_2 - \kd^-_2/2 , x_\perp \} ,\ldots )  \nn \\
\ata \hf \int  \frac{dy^-}{2 \pi} d u^-
\frac{e^{i y^-(x_A p^+ + l^+ )} e^{i u^-(x_B p^+ - l^+ )} }
{(k^+ - i\e)  (l^+ - i \e) } \nn \\
\ata \lc k_1 - \D/2 | \psibar(0, x_\perp) \g^+ \psi(y^-, x_\perp) | k_1 + \D/2 \rc
\lc k_2 + \D/2 | F_\sg^+  (-u^-/2, x_\perp )  F^{\sg +} (u^-/2, x_\perp)  | k_2 - \D/2 \rc . 
\label{nucleon_mat_x2}
\eea

\nt
In the above equation, $\D^+ = k^+ + l^+ - x_C p^+$, we continue to use $\D^+$ to 
save writing. 
%The remaining four $\h$-functions may be expressed in terms of 
%contour integrals as in Eq.~(\ref{theta}), 
%\[ 
%\int \frac{d z_1 d z_2 d z_3 d z_4 }{(2\pi i)^4} \frac{ e^{i z_1 (k_1 + \D/2)  + 
%
%i z_2 (k_1 - \D/2) + i z_3 (k_2 - \D/2) + i z_4 (k_2 + \D/2) }  }
%
%{ ( z_1 - i \e )  ( z_2 - i \e ) ( z_3 - i \e )  ( z_4 - i \e )}. 
%\]
%The four space variables $z_1,z_2,z_3,z_4$ may be absorbed into shifts of 
%$x_1,x_2,\kd_1,\kd_2$. Following that, the vanishing of the nuclear 
%wavefuntion at asymptotic $z_1, \ldots , z_4$ is used to complete the 
%respective contour integrations. 
The only unphysical variables remaining are the two momenta $k^+, l^+$. Again, 
following Ref.~\cite{Ji:1997gs}, we note that the the OFPD's also demonstrate 
minimal variation with $\D^+$ and as a result that dependence may be ignored.
Restricting attention solely on the part that depends on $k^+,l^+$ we 
obtain the integrals 
\[
\int \frac{ dk dl}{ (2 \pi i)^2} \frac{ e^{i k^+ (x_2^- - x_1^-)} e^{i l^+ (x_2^- - x_1^-)} 
e^{i l^+ (y^- - u^-)}}{ (k^+ - i\e)  (l^+ - i \e) } .
\]
This yields the obvious condition $\h (x_2^- - x_1^-) \h (x_2^- - x_1^- + y^- - u^-) $.
Within the picture adopted by the convolution model, $x_1,x_2$ are the locations of the
centers of the two nucleons that contain the struck quark and the soft gluon which 
scatters off the struck quark, where as $u,y$ are locations within a given 
nucleon. As a result in the case where the struck quark and the soft gluon 
originate within two separate nucleons (as is the case in this calculation), 
the second $\h$-function is identical to the 
first and the overall integration over $k^+,l^+$ simply yeilds the physical 
condition that the nucleon containing the struck quark be situated spatially 
ahead of the nucleon containing the soft gluon. The contour integration 
over the complex space of $k^+, l^+$ sets their values to zero, as a 
result  $\D^+ = x_C p^+$. By inspection of Eq.~(\ref{mod_factor}), one 
notes that the only values that may be assumed by $x_C$ are $0$ or 
$x_L$. As a result, we make the further approximation that $x_Cp^+ << k_1^+, k_2^+$
and neglect $x_C$ in the denominators of the top line of Eq.(\ref{nucleon_mat_x2}). 
Corrections to this approximation are either vanishing or suppressed by powers of 
$x_L^2$. Following Ref.~\cite{Osborne:2002st}, and given that the nuclear wavefunction 
is peaked for values of $k_1+ \simeq p^+$ we make the approximation of replacing the 
off-forward parton distributions with the regular distributions,

\bea 
\frac{p^+}{k_1^+} \int \frac{dy^-}{2\pi} e^{i y^- x_A p^+ } 
\lc k_1 - \D/2 | \psibar(0, x_\perp) \g^+ \psi(y^-, x_\perp) | k_1 + \D/2 \rc
= 2  Q \left(\frac{xp^+}{k_1^+},\frac{\D^+}{p^+}, \frac{M^2 - k_1^2}{M^2} \right) \simeq 2 f_q (x)
\eea

\bea
\int \frac{du^-}{2\pi} \frac{e^{i u^- x_B p^+ }}{k_2^+}  
\lc k_2 + \D/2 | F_\sg^+  (-u^-/2, x_\perp )  F^{\sg +} (u^-/2, x_\perp)  | k_2 - \D/2 \rc
= 2 G \left(\frac{xp^+}{k_2^+},-\frac{\D^+}{p^+}, \frac{M^2 - k_2^2}{M^2} \right) \simeq 2 f_g (x)
\eea

\nt
It should be pointed out that the above approximations for the parton distribution 
functions have only been demonstrated to hold in convolution with nuclear density 
distributions which are peaked around the mean values of the momenta of the nucleons 
(see Ref.~\cite{Osborne:2002st} for further details).

Incorporation of the above simplifications leads to the following expression for the 
representative matrix element,

\bea 
T &\simeq& \int_0^{P_{max}} \frac{d k_1^+ d k_2^+}{(2 \pi)^2} 
A^2 \int d x_1^- d x_2^- d \kd_1 d \kd_2  d^2 x_\perp \prod_{i=3}^{A-1} d^3 x_i  
e^{i k_1 \kd_1 } e^{ i k_2 \kd_2} e^{i x_C p^+ ( x_2^- - x_1^- )} \nn \\
\ata \Phi^* (\{ x_1^- + \kd^-_1/2, x_\perp \} , \{ x^-_2 + \kd^-_2/2 , x_\perp \} , \ldots ) 
\Phi ( \{ x_1^- - \kd^-_1/2 , x_\perp \}, \{ x^-_2 - \kd^-_2/2 , x_\perp \} ,\ldots )  \nn \\
\ata 2\pi  f_q(x_A) f_g(x_B) . 
\eea

\nt
In the above equation, $P_{max}$ is the maximum allowed momentum in the nucleus, in principle 
this should be weakly dependent on the total number of nucleons in the nucleus, we will 
ignore this dependence in this effort. The integrals 
\[
\int_0^{P_{max}} \frac{dk_n}{2\pi} e^{ik_n \kd_n} ,
\]
\nt
are sharply peaked around the point $ \kd _n = 0 $  with a height proportional to $P_{max}$. We thus 
replace these integrals with a saddle point approximation around the point $\kd_n =0$. All factors 
of the width of the integrals will be included within an overall unknown normalization constant $B$.
This approximation simplifies the product of the nuclear wave-function and its complex conjugate 
$
\Phi^* (\{ x_1^- + \kd^-_1/2, x_\perp \} , \{ x^-_2 + \kd^-_2/2 , x_\perp \} , \ldots ) 
\Phi ( \{ x_1^- - \kd^-_1/2 , x_\perp \}, \{ x^-_2 - \kd^-_2/2 , x_\perp \} ,\ldots ) \ra 
\Phi^* (\{ x_1^-, x_\perp \} , \{ x^-_2, x_\perp \} , \ldots ) 
\Phi ( \{ x_1^-, x_\perp \}, \{ x^-_2, x_\perp \} ,\ldots ) = \rho[(x_1,x_\perp),(x_2,x_\perp) ]
$.
This is the two nucleon correlation function in a nucleus. We make the last approximation of 
replacing this with a product of two single nucleon densities and an unknown constant $D$. \tie,

\bea
\ro(x_1,x_2) = D \ro(x_1) \rho(x_2).
\eea

\nt this leads to the final simplified expression for the representative matrix elements which 
has already appeared previously in Refs.~\cite{guowang,maj04e}, 

\bea
T = \int d x_1^- d x_2^- d^2 x_\perp C  \ro (x_1^-,x_\perp) \ro (x_2^-, x_\perp) 
e^{ix_C p^+(x_2^- -x_1^-)} \h (x_2^- - x_1^-) A^2 f_q(x_A) x_B f_g (x_B), \label{mod_factor_simple}
\eea

\nt
where all normalization constants including factors of $2\pi$ and $p^+$ have been absorbed into 
the dimensionful constant~$C$. 

Such a constant may not be determined from first principles 
in our approach, 
but should, in principle, depend on the kinematics of a given experiment. As our derivation 
has demonstrated, it also may have a mild variation on the nucleus chosen. However, 
we will assume that the overall constant be independent 
of the nucleus chosen. Ostensibly, the overall constant also has no dependence on the 
fragmentation of the escaping jet and thus may not depend on the momentum fractions of the 
detected hadrons $z_1,z_2$ nor on the number of hadrons detected. Once a data point 
for a choice of momentum fraction (fractions) of a detected hadron (hadrons) is described 
by tuning $C$, it is set for all other momentum fractions and nuclei that may be used in the given 
experimental kinematics. The constant has a strong dependence on $p^+$, however, as in most 
experimental setups the energy per nucleon in an accelerated nucleus is held fixed for 
different nuclei, there is no variation across nuclear targets for different total momenta $P_A^+$.

In the present section, we have introduced a number of approximations in the evaluation of the 
medium modification factor outlined in Eq.~(\ref{mod_factor}). The approximations have 
resulted in a very much simplfied expression for the modification factor. In the subsequent 
section we will evaluate the modification of the dihadron fragmentation functions using 
the expression from Eq.~(\ref{mod_factor_simple}). Such a computation will require a 
certain model of the nuclear density distribution $\ro(\vx)$. Calculations for 
two separate distributions (gaussian, hard-sphere) will be presented.

%%%%%%%%%%%%%%%%%%%%%%%%%%%%%%%%%%%%%%%%%%%%%%
%%%%%%%%%%%%%%%%%%%%%%%%%%%%%%%%%%%%%%%%%%%%%%
%%%%%%%%%%%%%%%%%%%%%%%%%%%%%%%%%%%%%%%%%%%%%%
%%%%%%%%%%%%%%%%%%%%%%%%%%%%%%%%%%%%%%%%%%%%%%
%%%%%%%%%%%%%%%%%%%%%%%%%%%%%%%%%%%%%%%%%%%%%%
%%%%%%%%%%%%%%%%%%%%%%%%%%%%%%%%%%%%%%%%%%%%%%

\section{Numerical Results}

%%%%%%%%%%%%%%%%%%%%%%%%%%%%%%%%%%%%%%%%%%%%%%
%%%%%%%%%%%%%%%%%%%%%%%%%%%%%%%%%%%%%%%%%%%%%%
%%%%%%%%%%%%%%%%%%%%%%%%%%%%%%%%%%%%%%%%%%%%%%
%%%%%%%%%%%%%%%%%%%%%%%%%%%%%%%%%%%%%%%%%%%%%%
%%%%%%%%%%%%%%%%%%%%%%%%%%%%%%%%%%%%%%%%%%%%%%
%%%%%%%%%%%%%%%%%%%%%%%%%%%%%%%%%%%%%%%%%%%%%%

The modification of single and double hadron fragmentation functions in nuclei 
has been studied experimentally by the HERMES experiment at 
DESY \cite{Airapetian:2000ks,Airapetian:2003mi,din04}. 
The modification is presented as a ratio of the fragmentation of a hard parton 
produced in the DIS off a large nucleus, versus a deuterium nucleus. Within 
the kinematics of the experiment, the DIS off the deuterium 
nucleus leads to minimal modification of the fragmentation process. Fragmentation of the 
jet produced in such a process is assumed to be identical to fragmentation in vacuum.

In the rest frame of the struck nucleus the experiment can measure both the forward momenta 
of the produced particles $p_1^-$,$p_2^-$, the forward momentum of the virtual photon $q^- = \nu$ as 
well as the virtuality of the photon $Q^2$.
The momentum fractions are thus given simply as the ratios $z_1 = p_1^-/\nu$ and $z_2 = p_2^-/\nu$.
The modification of the single fragmentation function at a momentum fraction $z$ 
is represented by the ratio of the multiplicity of 
hadrons $N(z)dz$ with a momentum fraction in the range from $z$ to $z+dz$ 
produced in DIS off a large nucleus versus that produced in DIS off a deuterium nucleus:
\bea
R_1(z) = \frac{N_A(z)}{N_D(z)}.
\eea 
Theoretically this ratio is equated with the ratio of the medium modified fragmentation function versus that in the 
vacuum, \tie,
\bea 
R_1(z) = \frac{\T{D}(z,\nu,Q^2,A) }{D(z,Q^2)}.
\eea
Mesurements for a Nitrogen (N) and Krypton (Kr) nucleus are presented as the square and 
circular points in Figs.~\ref{fig10b},\ref{fig12}.

The  medium modification of the associated hadron fragmentation function is obtained by measuring 
the number of events with at least two hadrons with momentum 
fractions $z_1,z_2$ ($N^{(2)}_A(z_1,z_2)$),
and the number of events with at least one hadron with momentum fraction $z$ ($N_A(z)$). 
As in the case of the single fragmentation function a double ratio is presented using similar measurements off deuterium, 

\bea
R_2(z_2) = \frac{\frac{ \sum_{z_1>0.5} N^{(2)}_A(z_1,z_2)  }{ \sum_{z>0.5} N_A (z) }}
{\frac{\sum_{z_1>0.5} N^{(2)}_D(z_1,z_2)}{ \sum_{z>0.5} N_D (z) }}.
\eea
\nt
This double ratio for different $z_2$ is plotted in 
Figs.~\ref{fig11} and \ref{fig13} as the square points for nitrogen 
and the circular points for Krypton.
In the limit of low multiplicity per event, and the assumption that there 
is minimal modification in Deuterium 
the above ratio is theoretically estimated as 

\bea
R_2(z_2) = \frac{\frac{ \int_{0.5}^{1-z_2} dz_1 \tilde{D}_q(z_1,z_2,\nu,Q^2,A) }{ \int_{0.5}^{1} d z 
\tilde{D}_q(z,\nu, Q^2 , A)  }}
{\frac{\int_{0.5}^{1-z_2} dz_1  D_q(z_1,z_2,Q^2)}{ \int_{0.5}^{1} d z D_q(z,Q^2) }}.
\eea

In the remainder of this section we focus on the evaluation of the medium 
modification factor and the medium modification of single and 
double fragmentation functions. 
In the preceeding section, a set of approximations to the medium modification factor of 
Eq.~(\ref{mod_factor}) were carried out.
There exist four terms in Eq.~(\ref{mod_factor}), each of which may be expressed in 
general as in Eq.~(\ref{general_term}). The generic term, under the approximations instituted 
in the previous section may be expressed as Eq.~(\ref{mod_factor_simple}). Substituting 
Eq.~(\ref{mod_factor_simple}) back into the complete expression for the modification 
factor Eq.~(\ref{mod_factor}) leads to the expression,

\bea
T^A_{q g} (x, x_L) &=& \int d x_1^- d x_2^- d^2 x_\perp C  \ro (x_1^-,x_\perp) \ro (x_2^-,d x_\perp) 
\h (x_2^- - x_1^-) A^2 \nn \\ 
\ata \bigg[ 
f_q(x+x_L) x_T f_g (x_T) \bigg( 1 - e^{-ix_L p^+ ( x_2^-  - x_1^- )} \bigg) 
+ f_q(x) x_L f_g (x_L) \bigg( 1 -  e^{ix_L p^+ ( x_2^-  - x_1^- )} \bigg) \bigg], 
\label{mod_factor_simple2}
\eea 
 
\nt
which contains all the four terms of Eq.~(\ref{mod_factor}). To proceed further, 
an expression for the nuclear density distribution has to be substituted in 
the above equation. Two different choices of such a distribution will be studied in 
this article: a Gaussian distribution, which essentially concentrates most of the nucleons 
towards the center of the nucleus and a hard sphere distribution where the nucleons 
are distributed evenly over the entire nuclear volume. No doubt, most nuclei lie 
somewhere in between these two extremes. We refrain from using a Woods-Saxon 
distribution  \cite{ws}, as it becomes analytically intractable. The hard-sphere
distribution is a good approximation of the Woods-Saxon distribution for large nuclei.
In the following we present results for each of the different cases in turn.

The computation of the modification factor will be followed by the calculation 
of the modification of the fragmentation functions in Eq.~(\ref{med_mod}) for the 
dihadron fragmentation function and for the single 
fragmentation function following Ref.~\cite{guowang}. As is 
clear from Eq.~(\ref{med_mod}), the evaluation of the medium modified fragmentation 
function also requires the input of a vacuum dihadron fragmentation function and vacuum 
single fragmentation functions. The vacuum dihadron fragmentation functions may not be 
predicted entirely with QCD but require a measurement at a given energy scale. In the 
absence of experimental data we follow Ref.~\cite{maj04d} and parametrize
two-particle correlations in the tuned Monte-Carlo event generator JETSET \cite{sjo95}. 
The vacuum single fragmentation functions are tabulated and parametrized by various collaborations. 
We choose the parametrization of Ref.~\cite{bin95}. This parametrization is also consistent
with single particle distribution from our chosen event generator.

%%%%%%%%%%%%%%%%%%%%%%%%%%%%%%%%%%%%%%%%%%%%%%
%%%%%%%%%%%%%%%%%%%%%%%%%%%%%%%%%%%%%%%%%%%%%%
%%%%%%%%%%%%%%%%%%%%%%%%%%%%%%%%%%%%%%%%%%%%%%

\subsection{Gaussian distribution}
%%%%%%%%%%%%%%%%%%%%%%%%%%%%%%%%%%%%%%%%%%%%%%
%%%%%%%%%%%%%%%%%%%%%%%%%%%%%%%%%%%%%%%%%%%%%%
%%%%%%%%%%%%%%%%%%%%%%%%%%%%%%%%%%%%%%%%%%%%%%

Imagine a large nucleus with $A$ nucleons at rest, at a  given time $t_0$. 
The mass of each of the nucleons is $M$, and the nuclear radius is 
$R_A$.  
The spatial distribution of nucleons is assumed to be that of a Gaussian, \tie,

\bea
\ro(y,x_\perp) =  \ro_0 \exp\llb - \frac{y^2 + x_\perp^2}{2 R_A^2} \lrb 
= \ro_0 \exp \llb - \frac{x_\perp^2}{2 R_A^2} \lrb \exp\llb - \frac{{y^-}^2}{2 R_A^2} \lrb ,
\eea 

\nt
where $\ro_0$ is the normalization factor, $y$ is the coordinate in the 
$3$-direction and $\vx_\perp$ is the two dimensional vector orthogonal to it. 
The lightcone vector is $y^- = t_0 - y$ at $t_0 = 0$. 
The density is normalised to unity, \tie,

\bea
\int d y^- d^2 x_\perp \ro_0 e^{ - \frac{x_\perp^2}{2 R_A^2  } } e^{- \frac{{y^-}^2}{2 R_A^2}} = 1  
\imp \rho_0 = \frac{1}{ (2 \pi)^{3/2} R_A^3}
\eea

We then boost the nucleus in the $y$-direction to a frame where its 
momentum $P_A = A[\sqrt{M^2 + p^2},0,0,p]$ is very large. For a large enough boost, 
we may approximate the light cone coodinates in the boosted frame $y_b^- = y^-/2\g$, 
where $\g = p^+ / M$.
The intergral of the density in the coordinates of the boosted frame becomes, 
(dropping the subscript $b$),

\bea
\int d y^- d^2 x_\perp 2 \g \ro(y^-,x_\perp) =
\int d y^- d^2 x_\perp 2 \g \ro_0 e^{ - \frac{x_\perp^2}{2 R_A^2  } }  
e^{- \frac{{y^-}^2}{2 (R_A/2\g)^2}} = 1 
\eea

\nt
The overall factor of $2\g$ may be absorbed into a redefinition of $\ro_0$. 
%We also introduce the lightcone radius $R_A^- = R_A/2\g $. 
This expression for the density is 
then substituted into Eq.~(\ref{mod_factor_simple2}). There exist two kinds of terms 
in Eq.~(\ref{mod_factor_simple2}): those with and without the phase factor 
$\exp [ \pm ix_L p^+ ( x_2^-  - x_1^- )]$. Summing the terms without the 
phase factor, one obtains,

\bea 
T^{A(1)}_{q g} (x, x_L) &=& \int d x_1^- d x_2^- d^2 x_\perp 2 C \frac{2\g }{(2\pi)^3  R_A^6}   
\h (x_2^- - x_1^-) A^2 \nn \\ 
\ata \ro_0 e^{ - \frac{x_\perp^2}{ R_A^2  } }  e^{- \frac{{x_1^-}^2  +  {x_2^-}^2 }{2 (R_A/2\g)^2}}
f_q(x+x_L) x_T f_g (x_T) \label{gauss_no_phase}
\eea

It is now trivial to carry out the Gaussian integrals over the position variables 
$x_\perp,x_1^-,x_2^-$ which yields factors of $2 \pi$, $\g$ and $R_A^4$. 
%The nuclear radius $R_A = (3 A M / 4 \pi \rho_{nucleon})^{1/3} $, where  $\ro_{nucleon}$ is the 
%fundamental nucleon matter density. 
Each power of $R_A$ yields a factor of $A^{1/3}$. 
Counting the power of $A$ and absorbing all other factors of $p^+$ and $2\pi$ 
into the overall normalization constant, we obtain the form as in Ref.~\cite{guowang}

\bea 
T^{A(1)}_{q g} (x, x_L) &=& \T{C} M R_A f_q^A (x + x_L) x_T f_g (x_T),
\eea 

\nt
where we have approximated the quark structure function in a nucleus to be simply $A$ times that 
in a nucleon $f_q^A (x + x_L) =  A f_q(x+x_L) $. 

The terms with the phase factor undergo a similar intergration, these 
%with the phase factor 
%absorbed into the Gaussian weight leaving an overall exponential damping factor' \tie
may be summed to obtain 
%\[
%\exp \llb- \frac{{x_1^-}^2  +  {x_2^-}^2 }{2 (R_A/2\g)^2}\lrb \exp [ \pm ix_L p^+ ( x_2^-  - x_1^- )]
%
%= \exp \llb- \frac{(x_1^- + x_2^-)^2 }{4 (R_A/2\g)^2}
%
% -\left( \frac{ (x_1^- - x_2^- ) }{2 (R_A/2\g)} - i x_L p^+ \frac{R_A}{2\g} \right) - 
%\frac{x_L^2}{x_A^2} \lrb.
%
%\]

\bea 
T^{A(1)}_{q g} (x, x_L) &=& - \int d x_1^- d x_2^- d^2 x_\perp 2 C \frac{2\g }{(2\pi)^3  R_A^6}   
\h (x_2^- - x_1^-) A^2 \nn \\ 
\ata \ro_0 e^{ - \frac{x_\perp^2}{ R_A^2  } }  e^{- \frac{{x_1^-}^2  +  {x_2^-}^2 }{2 (R_A/2\g)^2}}
f_q(x+x_L) x_T f_g (x_T) \cos[ x_L p^+ ( x_2^- - x_1^- )] \label{gauss_with_phase}
\eea

\nt
The part of the above exponentials that depend on the positions of the two nucleons may now be 
integrated out as before.
Adding the the contributions (both with and without the phase factor) 
results in the final expression for the medium modification 
function, in the Gaussian density approximation,

\bea 
T^A_{qg}(x,x_L) &=& \frac{\tilde{C}}{x_A} \bigg[ f_q^A(x) x_T f_g^N (x_T) + 
f_q^A(x) (x_L+x_T) f_g^N (x_L + x_T)  \bigg] (1 - e^{-x_L^2/x_A^2}),
\eea

\nt
where $x_A = 1/(MR_A)$. The above expression for the modification factor is then
used in Eq.~(\ref{mod_split}) to obtain the modified splitting functions and
the medium modified fragmentation functions in the Gaussian density 
approximation. There still remains the overall factor $\T{C}\sim C[xf_g^N(x)]$
which represents the correlation between the nucleons and the gluon distribution
at small $x$. We will consider this factor a fit parameter. It is determined for one 
measurement of the fragmentation function (single or double) at one value of momentum 
fraction of the hadrons in the DIS off one nucleus. We will use that value to
predict medium modification of both single and dihadron fragmentation functions.

\begin{figure}[htbp]
\begin{center}
  \epsfig{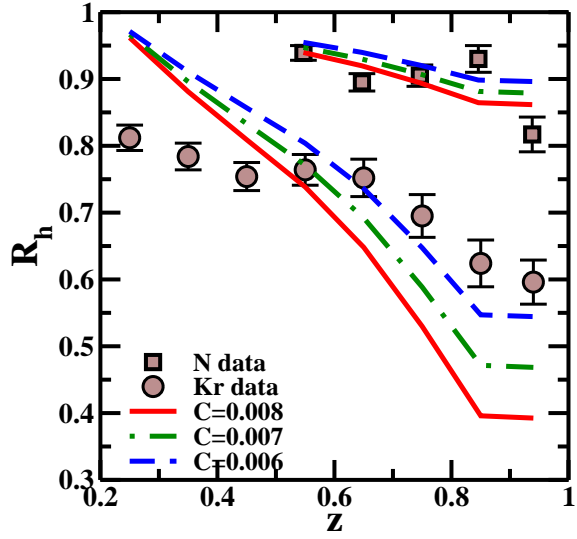}
%\hspace{0cm}
    \caption{ The medium modification of the single fragmentation functions compared to data from the 
HERMES collaboration. 
    A Gaussian distribution of nucleons in a nucleus is used. 
    The dotted lines indicate a 
    $\T{C}=0.006$, the dot-dashed lines a $\T{C}=0.007$ and the solid line a $\T{C} = 0.008$. }
    \label{fig10b}
  \end{center}
\end{figure}

The modifification of the single and double fragmentation functions
in the DIS off nuclei ($\T{D}(z),\T{D}(z_1,z_2) $) has been measured by 
the HERMES experiment at DESY. 
The modification for the single inclusive distribution 
is plotted in Fig.~\ref{fig10b} for Nitrogen (squared points) and
Krypton (circular points) targets . These are then divided by the identical 
measurement on the deuterium nucleus, assuming minimal modification of 
the fragmentation function in deuterium. One notes a continuing suppression 
with increasing momentum fraction $z$ and increasing nuclear size.
The suppression factor is expressed in our calculation as the ratio of 
the medium modified to the vacuum fragmentation functions. 
The value of $\T{C}$ is set by fitting the point at the lowest $z$ for Nitrogen
targets. Three different choices for $\T{C}=0.006,0.007,0.008$ are presented. 
Arguably the best fit is with $\T{C}=0.006$. 
The variation with increasing $z$ is then a prediction from 
Eq.~(\ref{med_mod}) and is shown as the solid line for the N nucleus. 
The calculation of the suppression factor in $Kr$ requires nothing more 
than a change of the number $A$ from 14 to 81. One notes good agreement with 
the data. However the agreement seems to deteriorate for smaller $z$.

\begin{figure}[htbp]
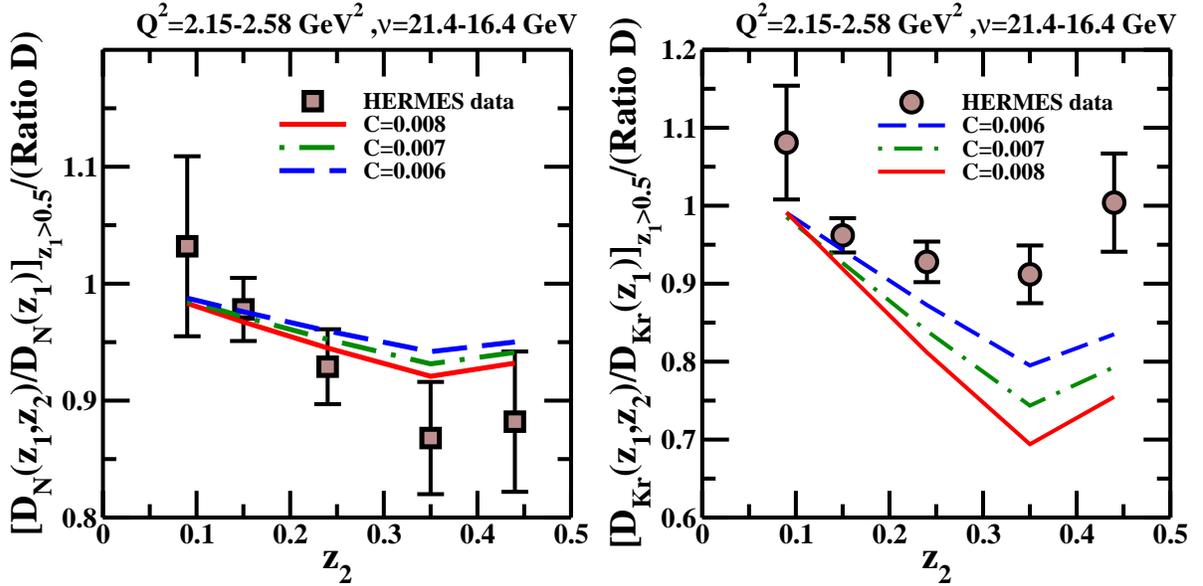

%\begin{center}
%  \epsfxsize 80mm
\hspace{-0cm}
  \epsfig{file=hermes_double_N_gauss.eps,scale=0.7}  
\hspace{0cm}
  \epsfig{file=hermes_double_Kr_gauss.eps,scale=0.7}  
     \caption{ Results of the medium modification of the quark fragmentation 
    function in cold nuclear medium in a N and Kr nucleus versus the momentum fraction of the associated 
    hadron. The momentum fraction of the trigger, is held above 0.5 and integrated over 
    all allowed values. A Gaussian distribution of nucleons in a nucleus is used. See text for details. }
    \label{fig11}
%  \end{center}
\end{figure}

The modification of the dihadron fragmentation function as a function of
momentum fraction for $N$ is presented in Fig.~\ref{fig10b}. It 
should be stressed once again that there are no free parameters in the 
calculation of the dihadron fragmentation function. The sole parameter 
$\T{C}$ was set by comparison to the medium 
modified single fragmentation function (Fig.~\ref{fig10b}). It turns out that within 
the choice of parameters the comparison with the $N$ nucleus is very good.
The modified dihadron fragmentation function for $Kr$ is shown 
in Fig.~(\ref{fig11}). One notes the agreement deteriorates for increasing $z$. 
The best fit to the data is obtained, once again with $\T{C} = 0.006$.

It would seem that the Gaussian approximation for the nuclear density distribution 
provides a good fit to the data for both single and double pion distributions in 
the case of $N$. The comparison is not so good in the case of $Kr$. 
However, it should be noted, that the Gaussian density approximation, which is 
close to the actual density distribution in $N$ is far from the actual 
nuclear density distribution in a larger nucleus such as $Kr$. In the next subsection,
we will use a hard sphere distribution which is a better approximation for large
nuclei.

%%%%%%%%%%%%%%%%%%%%%%%%%%%%%%%%%%%%%%%%%%%%%%
%%%%%%%%%%%%%%%%%%%%%%%%%%%%%%%%%%%%%%%%%%%%%%
%%%%%%%%%%%%%%%%%%%%%%%%%%%%%%%%%%%%%%%%%%%%%%

\subsection{Hard-sphere distribution}
%%%%%%%%%%%%%%%%%%%%%%%%%%%%%%%%%%%%%%%%%%%%%%
%%%%%%%%%%%%%%%%%%%%%%%%%%%%%%%%%%%%%%%%%%%%%%
%%%%%%%%%%%%%%%%%%%%%%%%%%%%%%%%%%%%%%%%%%%%%%

In this subsection nuclear density distribution will be 
assumed to be a hard-sphere,
\bea
\ro(y,x_\perp) =  \ro_0 \h (R_A - \sqrt{y^2 + x_\perp^2} ) . \label{rho_HS}
\eea

\nt
The normalisation,

\bea 
\int dy d^2 x_\perp \rho (y,x_\perp) = \ro_0 \frac{4 \pi}{3} R_A^3 = 1,
\eea

\nt
implies $\rho_0 =  3/(4 \pi R_A^3)$. As before, we now boost the 
nucleus to a frame where its forward momentum fraction is $Ap^+$. The 
nuclear density in terms of the variables $y^-, x_\perp$ is given as 
\bea
\ro (y^-,x_\perp) = 2 \g \ro_0 \h (R_A - \sqrt{  {y^-}^2 4 \g^2 + x_\perp^2} ).
\eea

We undertake the approximation that $(x_L+x_T)f_g^N (x_L + x_T) \simeq x_T f_g^N (x_T)$ for the 
gluon densities in Eq.~(\ref{mod_factor_simple2}). This leads to the simplified expression,

\bea 
T^{A}_{q g} (x, x_L) &=& \int d x_1^- d x_2^- d^2 x_\perp 2 C \frac{18 \g }{(4\pi)^2  R_A^6}   
\h (x_2^- - x_1^-) A^2 \ro_0  \h (R_A - \sqrt{  {x_1^-}^2 4 \g^2 + x_\perp^2} )\nn \\ 
\ata \ro_0  \h (R_A - \sqrt{  {x_2^-}^2 4 \g^2 + x_\perp^2} )
f_q(x+x_L) x_T f_g (x_T) [ 1 - \cos\{ x_L p^+ ( x_2^- - x_1^- ) \} ]. \label{HS_complete}
\eea

\nt
Integrating over the transverse coordinate, one obtains

\bea 
T^{A}_{q g} (x, x_L) &=& \int_{-R_A/2\g}^{R_A/2\g} d x_1^- \int_{x_1^-}^{R_A/2\g} d x_2^- 
[ R_A^2 - {(x_2^-  2 \g)}^2  ] \nn \\
\ata C A  \ro_0^2
f_q^A(x+x_L) x_T f_g (x_T) [ 1 - \cos\{ x_L p^+ ( x_2^- - x_1^- ) \} ]. 
\eea

\nt
In the above equation, one of the factors of $A$ has been absorbed into the definition of the 
quark distribution function in a nucleus $f_a^A(x)=A f_q(x)$. As in the case for the 
Gaussian density distribution, one may also absorb the gluon density 
$x_T f_g(x_T)$ into the over all normalization constant $\T{C}\sim C x_T f_g(x_T)$.
Integrating over $x_2$ and $x_1$ and absorbing all constants except for 
factors of mass number $A$, one obtains the final expresssion for the medium 
modification factor in the case of a hard-sphere distribution as 

\bea 
T^{A}_{q g} (x, x_L) &=& \T{C} \frac{f^A(x)}{x_A}  \Bigg[ \frac{1}{3} 
+ 4 \frac{x_A^3}{x_L^3} \sin \left(  \frac{x_L}{x_A} \right) - 8 \frac{x_A^4}{x_L^4} 
\left\{ 1 - \cos \left( \frac{x_L}{x_A} \right) \right\} \Bigg].
\eea

\nt
%While the enhancement with nuclear size is not as obvious as in the 
%case of the Gaussian distribution, the effect of the medium modification in the hard sphere 
%approximation is not dissimmilar. 

\begin{figure}[htbp]
\begin{center}
  \epsfig{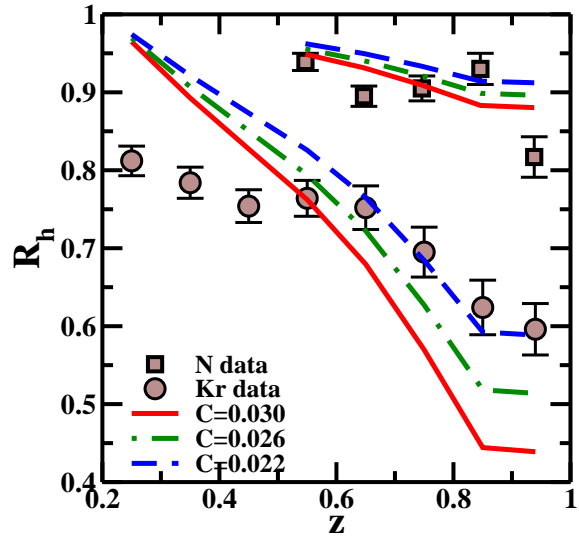}
%\hspace{0cm}
    \caption{ The medium modification of the single fragmentation functions compared to data from the 
HERMES collaboration. 
    A hard-sphere distribution of nucleons in a nucleus is used. The dotted lines indicate a
    $\T{C}=0.022$, the dot-dashed lines a $\T{C}=0.026$ and the solid line a $\T{C} = 0.030$. }
    \label{fig12}
  \end{center}
\end{figure}

Substituting the above result into the modified splitting functions in
Eq.~(\ref{mod_split}), one can calculate the medium modified 
fragmentation functions for a hard-sphere nuclear density distribution. There is the 
overall constant $\T{C}$ as in the case of the Gaussian density distribution. This 
is set by fitting to the data. Results for the medium modification of the single 
hadron fragmentation function in the hard sphere approximation are presented in 
Fig.~\ref{fig12} for three different values of the parameter $\T{C} = 0.022,0.024,0.030$
as compared with the identical data set as in Fig.~\ref{fig10b}. 
The data seem to prefer the value for $\T{C} = 0.022$. This fits both the 
lowest momentum fraction point in $N$ and provides a somewhat better description of 
the large $z$ region in Kr. In this sense a hard-sphere approximation seems to work 
better for the case of a large nucleus such as $Kr$. 

\begin{figure}[htbp]
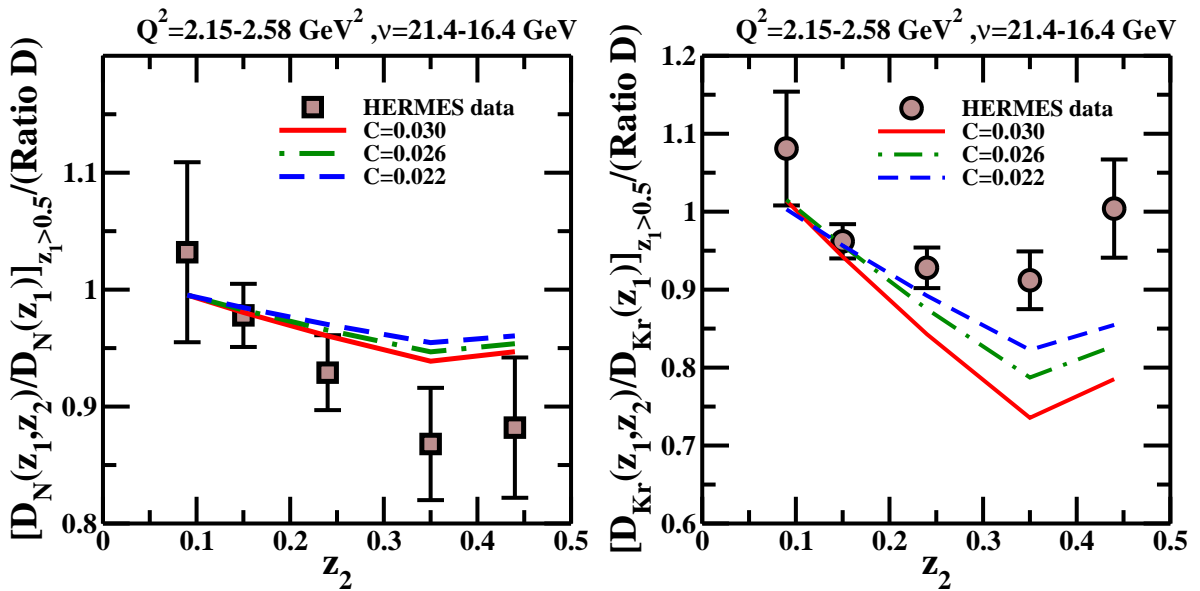

%\begin{center}
%  \epsfxsize 80mm
\hspace{-0cm}
  \epsfig{file=hermes_double_N_HS.eps,scale=0.7}  
\hspace{0.0cm}
  \epsfig{file=hermes_double_Kr_HS.eps,scale=0.7}  
     \caption{ Results of the medium modification of the quark fragmentation 
    function in cold nuclear medium in a N and Kr nucleus versus the momentum fraction of the associated 
    hadron. The momentum fraction of the trigger, is held above 0.5 and integrated over 
    all allowed values. A hard-sphere distribution of nucleons in a nucleus is used. See text for details. }
    \label{fig13}
%  \end{center}
\end{figure}

The modification of the dihadron fragmentation function as a function of the 
momentum fraction in the hard-sphere density approximation for $N$ is presented in 
Fig.~(\ref{fig13}). Once again, there are absolutely no free parameters in the 
calculation of the dihadron fragmentation function. The sole parameter 
$\T{C}$ was set by comparison to the medium modified single fragmentation 
function (Fig.~\ref{fig12}). It turns out that within 
the choice of parameters the comparison with the $N$ nucleus is not as 
good as the case for the Gaussian density distribution. This is to be 
expected as the Gaussian distribution is indeed closer to the actual 
density distribution in $N$. 

The modified dihadron fragmentation function for $Kr$ is 
presented in Fig.~(\ref{fig13}). One notes that while the agreement with the data still 
deteriorates for increasing $z$, there exits an overall quantitative improvement in the 
fit over that obtained from the Gaussian density distribution. 
The best fit to the data is obtained, once again with $\T{C} = 0.03$. It should be pointed 
out, in passing, that besides the slight improvement in the results for the case of the 
$Kr$ nucleus there remains no qualitative difference in the modification of the single and 
double fragmentation functions between the cases of a Gaussian distribution of nucleons and 
a hard sphere distribution.

\vspace{2cm}

The curves presented in Figs.~\ref{fig10b}-\ref{fig13} are the 
main results of this paper. They demonstrate that including next-to-leading twist corrections 
in the DIS off nuclei allow for a better understanding of the measured data.  However it is 
clear that one needs to go further to understand the behaviour displayed by the one and two 
particle distributions in the DIS off large nuclei such as $Kr$. For smaller nuclei such as $N$ 
the next-to-leading twist contributions seem to provide a satisfactory account of the 
experimental behaviour. 
%Further detailed discussion on this point, including a citation of 
%possible reasons for the disagreement as well as avenues for improvement, will be provided in 
%the subsequent section. 

%%%%%%%%%%%%%%%%%%%%%%%%%%%%%%%%%%%%%%%%%%%%%%
%%%%%%%%%%%%%%%%%%%%%%%%%%%%%%%%%%%%%%%%%%%%%%
%%%%%%%%%%%%%%%%%%%%%%%%%%%%%%%%%%%%%%%%%%%%%%
%%%%%%%%%%%%%%%%%%%%%%%%%%%%%%%%%%%%%%%%%%%%%%
%%%%%%%%%%%%%%%%%%%%%%%%%%%%%%%%%%%%%%%%%%%%%%
%%%%%%%%%%%%%%%%%%%%%%%%%%%%%%%%%%%%%%%%%%%%%%
\section{Discussions and Conclusions}

%%%%%%%%%%%%%%%%%%%%%%%%%%%%%%%%%%%%%%%%%%%%%%
%%%%%%%%%%%%%%%%%%%%%%%%%%%%%%%%%%%%%%%%%%%%%%
%%%%%%%%%%%%%%%%%%%%%%%%%%%%%%%%%%%%%%%%%%%%%%
%%%%%%%%%%%%%%%%%%%%%%%%%%%%%%%%%%%%%%%%%%%%%%
%%%%%%%%%%%%%%%%%%%%%%%%%%%%%%%%%%%%%%%%%%%%%%
%%%%%%%%%%%%%%%%%%%%%%%%%%%%%%%%%%%%%%%%%%%%%%

The focus of this paper has been on the medium modification of
dihadron fragmentation functions in the semi-inclusive
DIS off a large nucleus.
We have first generalized the formalism for the modification of the single 
fragmentation function in a dense medium to the modification of the dihadron 
fragmentation function. The modification arises from the inclusion of next-to-leading
twist contributions to the double differential inclusive cross section for
observing two hadrons within a jet produced via leptoproduction from a nucleus. 
Higher twist contributions are suppressed by powers of the hard scale $Q^2$ and
are thus ignored in the DIS off a nucleon target.
A class of these higher twist contributions are however enhanced by nuclear 
size $A^{1/3}$ \cite{lqs} and lead to 
the medium modification of both single and dihadron fragmentation functions. 

We have demonstrated (in Sec. ~II) that the medium modification of the dihadron 
fragmentation functions may be simply computed as a sum of convolutions involving 
medium modified splitting functions and vacuum dihadron fragmentation functions and 
and a qualitatively new contribution involving a new modified splitting 
function $\hat{P}_{q \ra qg}$ and a product of quark and gluon single 
hadron fragmentation functions [Eq.~(\ref{med_mod})]. Two of the modified 
splitting functions were shown to be identical to that in the case of the 
single fragmentation functions (\tie, $\D P_{q \ra qg}$ and $\D P_{q \ra gq}$). 
The new splitting function  $\D \hat{P}_{q \ra qg}$ was shown to be equivalent 
to $\D P_{q \ra qg}$ but without any virtual correction. Thus all modified 
splitting functions depend on the same medium modification 
factor $T_{qg}^A$ [Eq.~\ref{mod_factor}], identical to the case of the single 
fragmentation functions.

%Various approximations need to be made in the evaluation of the medium modification 
%factor. 
%It was demonstrated in Sec.~III that, due to color confinement, 
%the enhancement with nuclear size is only 
%restricted to matrix elements where the struck quark and the soft gluon 
%which scatters off it are contained in separate nucleons. 

These higher-twist contributions to the semi-inclusive DIS cross section
involves multiple scattering and gluon bremsstrahlung encountered by the 
struck quark and includes the well known Landau-Pomeranchuck-Migdal (LPM) 
interference effect. This interference effect leads to the appearance of the off-forward 
parton distribution functions (OFPD) \cite{Ji:1996ek} in the nuclear modification 
factor $T_{qg}^A$. Based on phenomenological 
models of OFPD's we approximated these by the regular forward parton distributions. 
The final result depends on the density distribution of the nucleons in the nucleus 
[Eq.~(\ref{mod_factor_simple2})]. Results for two different distributions were 
presented: Gaussian (Figs.~\ref{fig10b},\ref{fig11}) and hard-sphere 
(Figs.~\ref{fig12},\ref{fig13}). The two distributions demonstrated qualitatively 
similar effects on both single and double inclusive spectra.

The evaluation of the modification factor involved an overall normalization 
constant which represents the correlation between the nucleon containing the struck quark 
and the nucleon containing the soft gluon which scatter off the struck quark.
This was set by fitting to the overall experimental data on single inclusive distribution 
for DIS off $N$. The computation of the modification of the single fragmentation 
function on $Kr$ and the dihadron fragmentation function on both $N$ and $Kr$ thus 
involves no free parameters. The comparison of the theoretical prediction for the medium modified 
fragmentation functions (both single and double) with the experimental data for 
DIS off a $N$ nucleus is very good. The comparion for the case of the $Kr$ nucleus is 
somewhat satisfactory. A possible cause for this may lie in the inclusion of only 
the twist four contributions. These are contributions proportional to $A^{1/3}/Q^2$ 
that are suppressed by a power of $Q^2$ yet enhanced by a factor of $A^{1/3}$.
Physically this means that the struck quark may undergo at most two more scatterings 
off soft gluons prior to exiting the nucleus and fragmenting. 
Further scatterings necessarily involve higher twist parton correlation functions 
and require the inclusion of further powers of $A^{1/3}/Q^2$. While, such contributions 
may be unimportant in the case of $N$, they may become necessary for the 
case of $Kr$ and heavier nuclei. A systematic inclusion 
of all orders of $A^{1/3}/Q^2$ in the modification of the fragmentation process will 
involve a far more complicated calculation and is beyond the scope of this paper.

%%%%%%%%%%%%%%%%%%%%%%%%%%%%%%%%%%%%%%%%%%%%%%
%%%%%%%%%%%%%%%%%%%%%%%%%%%%%%%%%%%%%%%%%%%%%%
%%%%%%%%%%%%%%%%%%%%%%%%%%%%%%%%%%%%%%%%%%%%%%
%%%%%%%%%%%%%%%%%%%%%%%%%%%%%%%%%%%%%%%%%%%%%%
%%%%%%%%%%%%%%%%%%%%%%%%%%%%%%%%%%%%%%%%%%%%%%
%%%%%%%%%%%%%%%%%%%%%%%%%%%%%%%%%%%%%%%%%%%%%%

\section{Acknowledgement}

%%%%%%%%%%%%%%%%%%%%%%%%%%%%%%%%%%%%%%%%%%%%%%
%%%%%%%%%%%%%%%%%%%%%%%%%%%%%%%%%%%%%%%%%%%%%%
%%%%%%%%%%%%%%%%%%%%%%%%%%%%%%%%%%%%%%%%%%%%%%
%%%%%%%%%%%%%%%%%%%%%%%%%%%%%%%%%%%%%%%%%%%%%%
%%%%%%%%%%%%%%%%%%%%%%%%%%%%%%%%%%%%%%%%%%%%%%

The authors thank  R. Fries,  M. Gyulassy, B. M\"{u}ller, J. Qiu, 
and E. Wang  for helpful discussions. 
This work was supported in part by 
%the Natural Sciences and 
%Engineering Research
%Council of Canada, 
%in part by the Fonds Nature et Technologies 
%of Quebec, 
%and in part 
by the Director, Office of Science, Office of High Energy and Nuclear Physics, 
Division of Nuclear Physics, and by the Office of Basic Energy
Sciences, Division of Nuclear Sciences, of the U.S. Department of Energy 
under Contract No. DE-AC03-76SF00098 and No. DE-FG02-05ER41367.

\end{document}